\documentclass[pra,twocolumn]{revtex4-1}

\usepackage[utf8]{inputenc}
\usepackage[english]{babel}
\usepackage[T1]{fontenc}
\usepackage{lmodern}

\usepackage{graphicx}
\usepackage{amsmath}
\usepackage{amsfonts}
\usepackage{amssymb}
\usepackage{amsthm}

\usepackage{microtype}

\usepackage[breaklinks=true,colorlinks]{hyperref}

\newcommand{\realp}{\operatorname{Re}}

\newcommand{\Ekin}{\left\langle E_{\mathrm{kin}} \right\rangle}
\newcommand{\Intensity}{ \left< \left| \alpha \right|^2\right>}
\newcommand{\bunch}{\mathcal{B}}

\begin{document}

\title{Optomechanical cooling and self-trapping of low field seeking point-like particles}

\author{Arthur Jungkind}
\email{arthur.jungkind@uibk.ac.at}
\author{Wolfgang Niedenzu}
\author{Helmut Ritsch}
 
\affiliation{Institut f\"ur Theoretische Physik, Universit\"at Innsbruck, Technikerstra{\ss}e~21a, A-6020~Innsbruck, Austria}

\date{\today}

\begin{abstract}
  Atoms in spatially dependent light fields are attracted to local intensity maxima or minima depending on the sign of the frequency difference between the light and the atomic resonance. For light fields confined in open high-Q optical resonators the backaction of the atoms onto the light field generates dissipative dynamic opto-mechanical potentials, which can be used to cool and trap the atoms. Extending the conventional case of high field seekers to the regime of blue atom-field detuning, where the particles are low field seeking, we show that inherent nonlinear atom field dynamics still can be tailored to cool and trap near zero field intensity. Studying field intensity, particle localization and kinetic energy for cavity driving or pumping the particle from the side, we identify optimal parameter regimes, where sub-Doppler cooling comes with trapping and minimal atomic saturation.
\end{abstract}

\maketitle

\section{\label{sec:Introduction}Introduction}
Motional cooling of atomic gases by help of laser light has become one of the most fruitful areas of AMO physics in the past decades and allowed to reach the lowest temperatures and motional precision control of particles available in physics. While routinely relying on specific atomic level schemes and atomic spontaneous emission for entropy dissipation, it was already noted a long time ago \cite{Horak1997cavity}, that, in principle, any point like particle with an optical dipole moment can be cooled and trapped in a laser driven optical resonator. Here cavity decay replaces spontaneous emission as entropy sink and the temperature limit is only set by the optical cavity linewidth \cite{domokos2003mechanical}. For atomic transitions with sub-recoil frequency linewidth even much colder temperatures are predicted \cite{jager2017semiclassical}.

While cavity cooling and trapping was soon experimentally implemented for atoms and ions \cite{maunz2004cavity,murr2006three, schleier2011optomechanical}, it took much more time to apply it to nano-particles~\cite{kiesel2013cavity, millen2015cavity,stickler2016rotranslational,fonseca2016nonlinear}. It was only last year that experimental technology was so strongly improved that it is now very close to reaching the quantum ground state \cite{windey2018cavity, magrini2018near}. So far, still no convincing implementation was demonstrated for molecules, where ro-vibrational heating counteract the cooling process. Recently, substantial experimental evidence for cavity cooling towards degeneracy was found \cite{wolke2012cavity}. 

Conventional cavity cooling works with high field seeking particles so that the cooling is accompanied by optical dipole trapping at the cavity field anti-nodes. While this is the most straightforward and easy to implement combination it often suffers from the fact that the particles finally are kept at intensity maxima, where they experience internal heating and diffusion. This is particularly problematic for particles with much internal structure as molecules or nano-particles in vacuum. 
Here we compare this cooling to the new parameter regime of low field seeking particles. In principle these can be trapped at field intensity minima or even zeros, while still using cavity decay as a dissipation channel to cool and confine their motion. In an ideal case the particle would be kept close to a field node at close to zero intensity. Only when leaving the node area a stronger field would build up in the cavity to push the particle back. As in the red detuned case the two possible scenarios of longitudinal and transverse pumping can be considered. While the first case is at least conceptually simpler, the latter one is associated with spatial self-ordering and scales more favorable for larger particle numbers. Note that in any case we need the cavity to be blue shifted with respect to the laser to allow for kinetic energy extraction via cavity enhanced emission \cite{gangl20003d}.   

Note that recently an alternative scheme to trap at low intensities in the red detuned regime, called SIDA, was studied, which uses the nonlinearity of a very strong atom-field coupling \cite{neumeier2018reaching}. While trapping is predicted to work very well here, the cooling properties and steady state kinetic energies have yet to be determined in detail.   

\begin{figure}
\centering
\includegraphics[width=.7\linewidth]{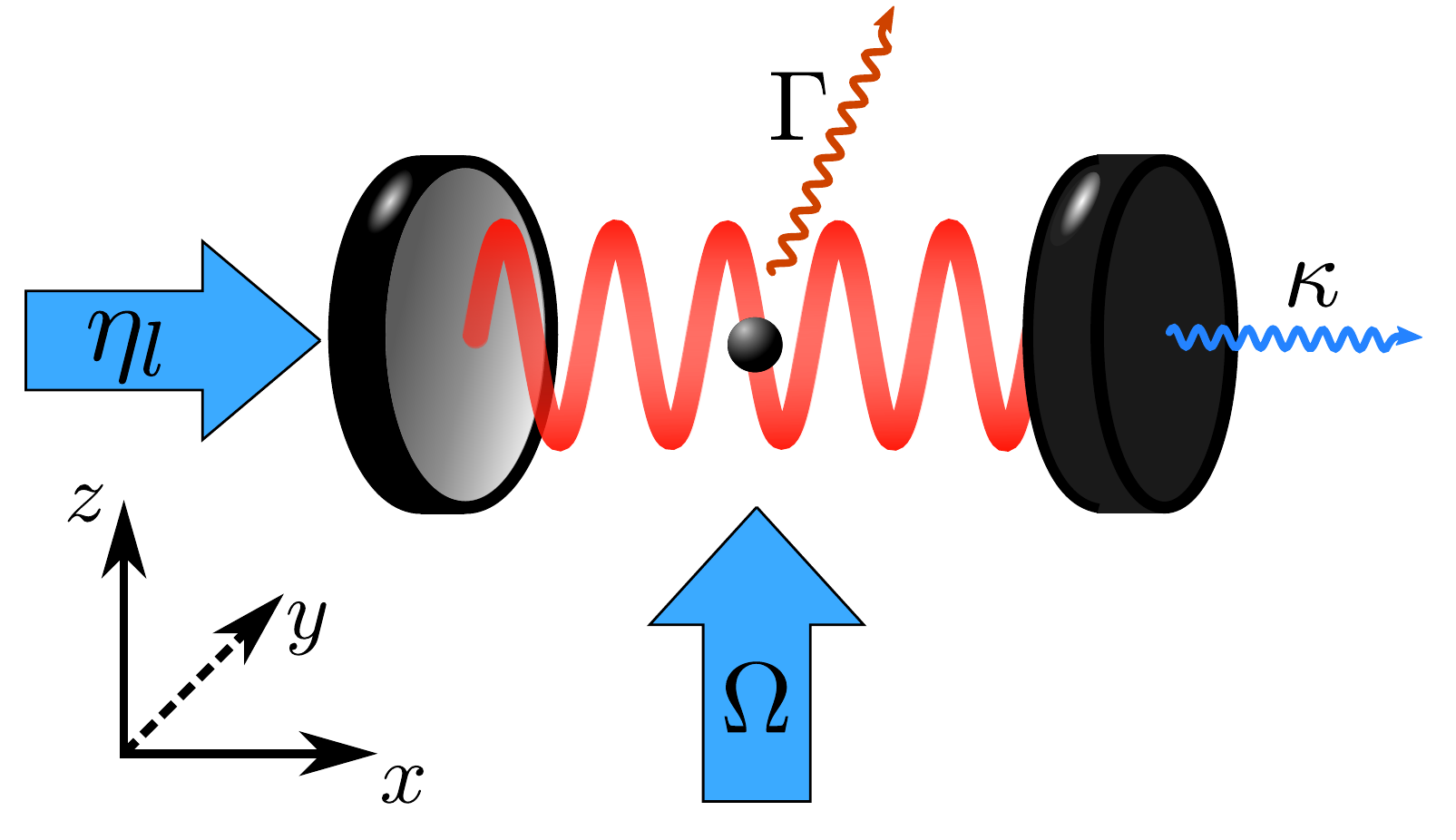}
\caption{Schematic drawing of a cavity cooling setup. The particle can spontaneously emit photons at rate $\Gamma$ and the cavity dissipation rate is $\kappa$. The system is driven either through a mirror with laser amplitude $\eta_{\mathrm{l}}$ or transversely from the side with the atomic Rabi frequency $\Omega$.}
\label{fig_setup}
\end{figure}

This work is organized as follows: we first introduce the model and review the original trapping and cooling predictions based on a linear semi-classical analytic model, which we compare to numerical simulations of a stochastic extension of these semi-classical equations including spontaneous emission and cavity losses. For the favourable parameter ranges in the blue detuned regime, we then analyze the optimal conditions to achieve low kinetic energies combined with localization in areas of close to zero intensity.

\section{\label{sec:Model}Model}

We consider a two-level atom with transition frequency $\omega_\mathrm{a}$ moving along the axis of a linear cavity (Fig.~\ref{fig_setup}). The atom interacts with a single standing wave cavity field mode of frequency $\omega_\mathrm{c}$ via the standard Jaynes--Cummings interaction~\cite{wallsbook}. We consider either (longitudinal) pump via one of the mirrors at amplitude $\eta_{\mathrm{l}}$ or a (transverse) laser drive onto the atom with Rabi frequency $\Omega$. The atom then scatters light from the pump laser into the resonator with a position-dependent phase at an effective pump strength $\eta_\mathrm{eff}$~\cite{ritsch2013cold}. The frequency $\omega_{\mathrm{p}}$ of the driving laser (pump) is detuned by $\Delta_{\mathrm{a}}=\omega_{\mathrm{p}}-\omega_{\mathrm{a}}$ and $\Delta_{\mathrm{c}}=\omega_{\mathrm{p}}-\omega_{\mathrm{c}}$ from the atomic and cavity resonances, respectively.

\par

In the standard dipole and rotating-wave approximations~\cite{wallsbook}, the Hamiltonian in a reference frame rotating with the pump frequency $\omega_{\mathrm{p}}$ reads~\cite{ritsch2013cold}
\begin{multline}
  H=\frac{p^2}{2m}-\hbar\Delta_{\mathrm{a}}\sigma_{+}\sigma_{-}-\hbar\Delta_{\mathrm{c}}a^{\dagger}a-i\hbar g\left(x\right)\left(\sigma_{+}a-a^{\dagger}\sigma_{-}\right) \\
   -i\hbar\eta_{\mathrm{l}}\left(a-a^{\dagger}\right)-i\hbar\Omega\left(\sigma_{+}-\sigma_{-}\right).
\label{Eq.TotalHamiltonian}
\end{multline}
Here $m$ is the mass of the atom, $a^{\dagger}$ and $a$ are the photonic creation and annihilation operators and $\sigma_{+}$ and $\sigma_{-}$ are the atomic excitation and de-excitation operators, respectively. The position-dependent coupling between the light field and the particle is $g\left(x\right)=g_{0}\cos\left(kx\right)$, where $k=\omega_\mathrm{c}/c$, $g_{0}=d\sqrt{\hbar\omega_{\mathrm{c}}/2\varepsilon_{0}\mathcal{V}}$ is the coupling parameter, $d$ denotes the atomic dipole moment and $\mathcal{V}$ is the cavity mode volume.

\par

The atom-cavity system is coupled to the surrounding electromagnetic vacuum, which gives rise to spontaneous emission of the atom at rate $\Gamma$ and cavity decay at rate $\kappa$. The non-unitary time evolution of the reduced density operator of the atom-cavity system is then given by the master equation~\cite{gardinerbook,wallsbook}
\begin{equation}
\dot{\rho}=-\frac{i}{\hbar}\left[H,\rho\right]+\mathcal{L}_{\mathrm{c}} \rho+\mathcal{L}_{\mathrm{a}}\rho,
\label{Eq.Mastereq}
\end{equation}
with the Liouville operators~\cite{ritsch2013cold}
\begin{subequations}
  \begin{align}
    \mathcal{L}_{\mathrm{c}}\rho &=\kappa\left(2a\rho a^{\dagger} -a^{\dagger} a\rho -\rho a^{\dagger}a\right), \label{Eq.LiouvilleA}\\
    \mathcal{L}_{\mathrm{a}}\rho&=2\Gamma\int d^2\textbf{u}N\left(\textbf{u}\right)\sigma_{-} e^{-i\textbf{u}x}\rho e^{i\textbf{u}x}\sigma_{+}\nonumber\\&\quad-\Gamma\left(\sigma_{+} \sigma_{-}\rho + \rho \sigma_{+} \sigma_{-} \right),\label{Eq.LiouvilleB}
  \end{align}
\end{subequations}
where $\textbf{u}$ is the direction of the emitted photon and $N(\mathbf{u})$ its spatial distribution.

\subsection{\label{sec:Master}Effective master equation}

For either sufficiently large detuning $\left|\Delta_{\mathrm{a}}\right|$ or spontaneous emission rate $\Gamma$ the internal atomic dynamics adiabatically follows the slower degrees of freedom. For low atomic saturation per photon,
\begin{equation}
  s=\frac{g_{0}^{2}}{\Delta_{\mathrm{a}}^{2}+\Gamma^{2}},
\end{equation}
the atom is, on average, mainly in its ground state and we can adiabatically eliminate the excited state. Substituting the steady-state atomic polarization
\begin{align}
\sigma^{\mathrm{st}}_{-}=-\frac{g\left(x\right)a+\Omega}{-i\Delta_{\mathrm{a}}+\Gamma}
\label{Eq.SteadySigma}
\end{align}
into the master equation~\eqref{Eq.Mastereq} yields~\cite{maschler2005cold}
\begin{subequations}
  \begin{multline}
    H_{\mathrm{eff}}=\frac{p^{2}}{2m}-\hbar\Delta_{\mathrm{c}}a^{\dagger}a+\hbar U_{0}\cos^{2}\left(kx\right)a^{\dagger}a-i\hbar\eta_{\mathrm{l}}\left(a-a^{\dagger}\right)\\+\hbar\eta_{\mathrm{eff}}\cos\left(kx\right)\left(a+a^{\dagger}\right),
    \label{Eq.EffHamil}
  \end{multline}
  where $\eta_{\mathrm{eff}}=\Omega g_{0}\Delta_{\mathrm{a}}/\left(\Delta_{\mathrm{a}}^2+\Gamma^2 \right)$ is the effective transverse pump rate, and
  \begin{equation}
     \mathcal{L}_{\mathrm{eff}}\rho=\left[\kappa+\Gamma_{0}\cos^{2}\left(kx\right)\right]\left(2a\rho a^{\dagger}-a^{\dagger}a\rho-\rho a^{\dagger}a\right).
     \label{Eq.EffLiou}
 \end{equation}
\end{subequations}
Here
\begin{subequations}\label{Eq.Dispersiveshift}
  \begin{align}
    U_{0}&=\frac{g_{0}^{2}\Delta_{\mathrm{a}}}{\Delta_{\mathrm{a}}^{2}+\Gamma^{2}}\label{Eq.U0}\\
    \Gamma_{0}&=\frac{g_{0}^{2}\Gamma}{\Delta_{\mathrm{a}}^{2}+\Gamma^{2}},\label{Eq.Gamma0}
\end{align}
\end{subequations}
describe the dispersive and absorptive effects of the atoms on the cavity-field mode~\cite{maschler2005cold}.

We specifically target the case of $\omega_{\mathrm{c}} \geq \omega_{\mathrm{p}} \geq \omega_{\mathrm{a}} $ to induce kinetic energy extraction via cavity scattering on the one hand as well as have low field seeking particles to localize them near intensity minima on the other hand. The repulsive character of light field maxima arises at a blue atomic detuning, i.e., $\Delta_{\mathrm{a}}>0$, which introduces a positive AC Stark-shift on the atom, such that it is repelled from high intensity regions. Note that transverse confinement of the particle can still be implemented via transverse higher order modes. This method was successfully implemented in one of the first cavity cooling experiments at MPQ in Munich and guaranteed to keep the particle sufficiently long in the resonator at low local intensity~\cite{puppe2007trapping}.   

\subsection{\label{sec:Semiclassical}Semi-classical approximation}

At not too low temperatures, where the coherence length of the atomic center-of-mass wave function is small compared to the cavity field wavelength, the momentum $p$ and position $x$ of the particle can be expressed by classical variables. Similarly, for sufficiently high photon number the coherent intra-cavity field can be approximated by a classical field with a complex amplitude $\alpha$. The master equation involving the effective Hamiltonian~\eqref{Eq.EffHamil} and the effective Liouvillian~\eqref{Eq.EffLiou} may then be mapped onto the following system of stochastic differential equations~\cite{domokos2003mechanical},
\begin{subequations}
\begin{align}
\dot{\alpha}&=\left\{ -\left[\kappa+\Gamma_{0}\cos^{2}\left(kx\right)\right]+i\left[\Delta_{\mathrm{c}} -U_{0}\cos^{2}\left(kx\right)\right]\right\} \alpha\nonumber\\
&\quad+\eta_{\mathrm{l}}-i\eta_{\mathrm{eff}}\cos\left(kx\right)+\xi_{\alpha}\label{Eq.Alpha}\\
\dot{p}&=-\frac{d}{dx}V\left(x\right)+\xi_{\mathrm{p}}\label{Eq.Momentum}\\
\dot{x}&=p/m,\label{Eq.Position}
\end{align}
\label{Eq.CoupledDGL}
\end{subequations}
with the intracavity trapping potential
\begin{align}
V\left(x\right)&=\hbar U_{0} \left|\alpha\right|^2\cos^2\left(kx\right)+2\hbar\eta_{\mathrm{eff}}\realp(\alpha)\cos\left(kx\right) \label{Eq.Potential}
\end{align}
and the Langevin noise terms $\xi_{\mathrm{\alpha}}$ and $\xi_{\mathrm{p}}$ originating from the vacuum field input and spontaneous emission of the atom~\cite{domokos2003mechanical}. Their correlation functions are specified in Appendix~\ref{app_noise}.

\subsection{Cavity cooling and heating}

\begin{figure}
\centering
\includegraphics[width=.49\linewidth]{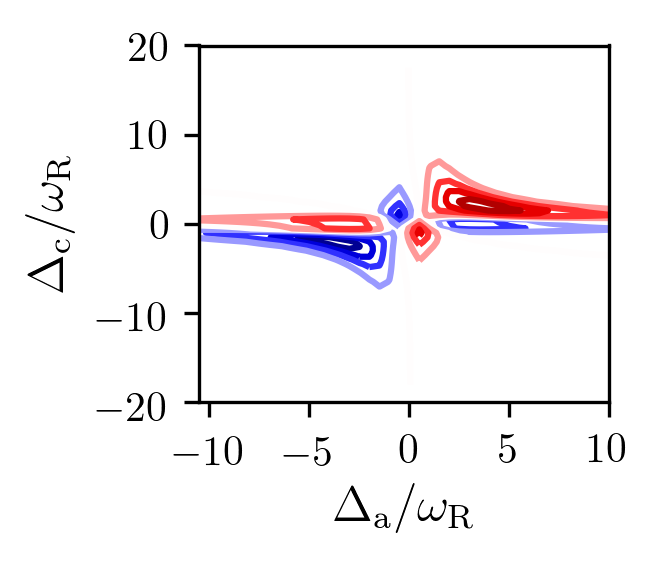}
\includegraphics[width=.49\linewidth]{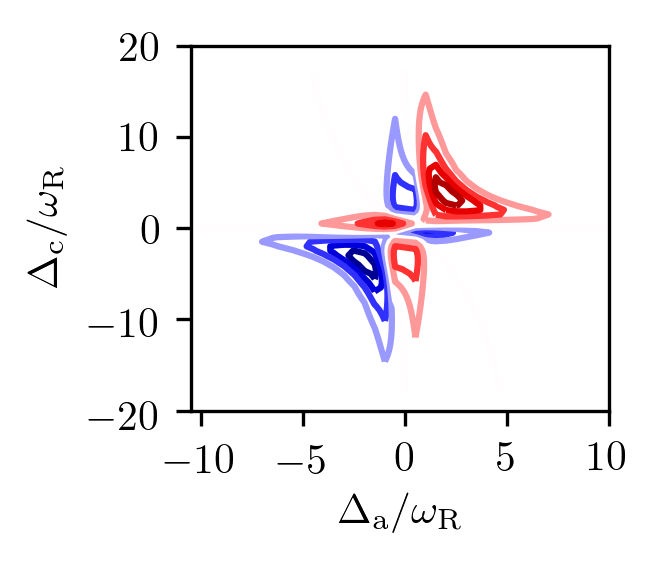}
\caption{Linear friction coefficient $f_{1}$ of a particle moving along the cavity axis in a cavity pump (left) and an atom pump (right) geometry as function of cavity detuning $\Delta_{\mathrm{c}}$ and atomic detuning $\Delta_{\mathrm{a}}$ averaged over one optical period. We chose $\eta_{\mathrm{l}}=1\,\omega_{\mathrm{R}}=\eta_{\mathrm{eff}}$, $\kappa=1\,\omega_{\mathrm{R}}$, $\Gamma=1\,\omega_{\mathrm{R}}$ and $g_0 = 3\,\omega_{\mathrm{R}}$. We find several regions of cooling (blue lines) and heating (red lines).}
\label{figfric}
\end{figure}

For a moving atom the field evolution lags behind the stationary values at given position thus yielding a velocity dependent force. A linear expansion of $a$ and $\sigma$ in slow velocities can be used to extract the linear friction coefficient
\begin{align}
f=f_{\mathrm{at}}+f_{\mathrm{ca}},
\end{align}
consisting of a Doppler cooling coefficient $f_{\mathrm{at}}$ arising from the inner dynamics of the atom and a cavity cooling coefficient $f_{\mathrm{ca}}$ arising from the cavity dynamics as described in \cite{hechenblaikner1998cooling} for cavity pump geometry and in \cite{domokos2002dissipative} for atom pump. The friction coefficient shows a rich structure as function of cavity and atomic frequency with respect to the pump laser. An average over one optical period is shown in Fig.\,\ref{figfric}, where typical cooling and heating regions are indicated by the blue and red solid lines. Note that for narrow atomic resonances and large atom-pump detunings the Doppler cooling plays only a minor role at low velocities. Hence, in part of the later simulations it is left out.\\

We see that for the case of red detuning with respect to atom and cavity one finds a region of strong friction which coincides with a maximum of the photon number in the cavity. Hence, the particles are drawn to the intensity maximum and cooled. This is the typical operation regime of cavity cooling and very well studied theoretically and experimentally.  
However, a closer look also reveals a cooling region for blue detuned pump light with respect to the atomic transition, i.e. $\Delta_{\mathrm{a}} > 0$. In this case the particle is low field seeking and potentially trapped at positions of minimal intensity resulting in low perturbation and diffusion. 

Similar to the friction coefficient $f_{1}$ also the momentum diffusion coefficient $D$ can be approximated in the close to zero velocity limit. As known from the famous Einstein relation we can thus estimate the stationary temperature via $k_{\mathrm{B}} T = D / f $ which is predicted to have a lower bound $k_{\mathrm{B}} T_{\mathrm{min}} \approx \hbar\kappa/2$ \cite{domokos2003mechanical}. The temperature is related to the kinetic energy via 
\begin{align}
    \Ekin = \frac{k_{\mathrm{B}} T}{2},
\end{align}
with $ E_{\mathrm{kin}} = p^2/2m$.

In the following numerical simulations we will check these previous predictions in a full dynamic model including spontaneous emission and cavity fluctuations. Here we centrally concentrate on the case of blue detuning - low field seeking particles - and the question how well particles can be trapped at minimal or even zero field positions and how low the corresponding temperature can get.

\section{\label{sec:Results}Numerical simulation of the system dynamics}

We investigate the important characteristics of our system by scanning of the corresponding parameter ranges. In correspondence to an experimental set-up the cavity decay rate as well as spontaneous emission rate are chosen to be fixed, as  $\kappa=40\,\omega_{\mathrm{R}}$ and $\Gamma=1\,\omega_{\mathrm{R}}$, with $\omega_{\mathrm{R}}=\hbar k^2/2m$ being the recoil frequency. Moreover, we chose $g_0 = 80\,\omega_{\mathrm{R}}$. The parameters $\Delta_{\mathrm{a}}$, $\Delta_{\mathrm{c}}$, $\eta_{\mathrm{l}}$ and $\eta_{\mathrm{eff}}$ are accessible by tuning the pump frequency and increasing/decreasing the pump intensity, or changing the cavity volume. A numerical integration of Eqns.\,\eqref{Eq.CoupledDGL} was performed for a variety of these parameters, in the case where the initial kinetic energy of the particle is much larger than the intra-cavity potential depth. The initial position of the particle for different simulation runs inside the cavity is normally distributed around the center of the unit cell. Equally the initial momentum is drawn from a normal distribution around zero, such that we can define an initial kinetic temperature $k_{\mathrm{B}}T_{0} = 15 \,\hbar\kappa$. The initial cavity intensity is zero.

Each illustrated data point was obtained by averaging the numerical data over the last $20\,\omega_{\mathrm{R}}^{-1}$ of the steady state and over $2000$ trajectories. In the following section we want to look at the properties of the two systems in different parameter regimes and compare them with each other.

\subsection{\label{sec:Long}Longitudinal cavity pump}

First let us consider a linear standing wave cavity that is pumped through one of the cavity mirrors ($\eta_{\mathrm{l}} \neq 0$ and $\eta_{\mathrm{eff}}=0$).

\subsubsection{\label{sec:longDaDc}Atom detuning vs.\ cavity detuning - scan}

\begin{figure*}[t]
\centering
\includegraphics[width=.32\linewidth]{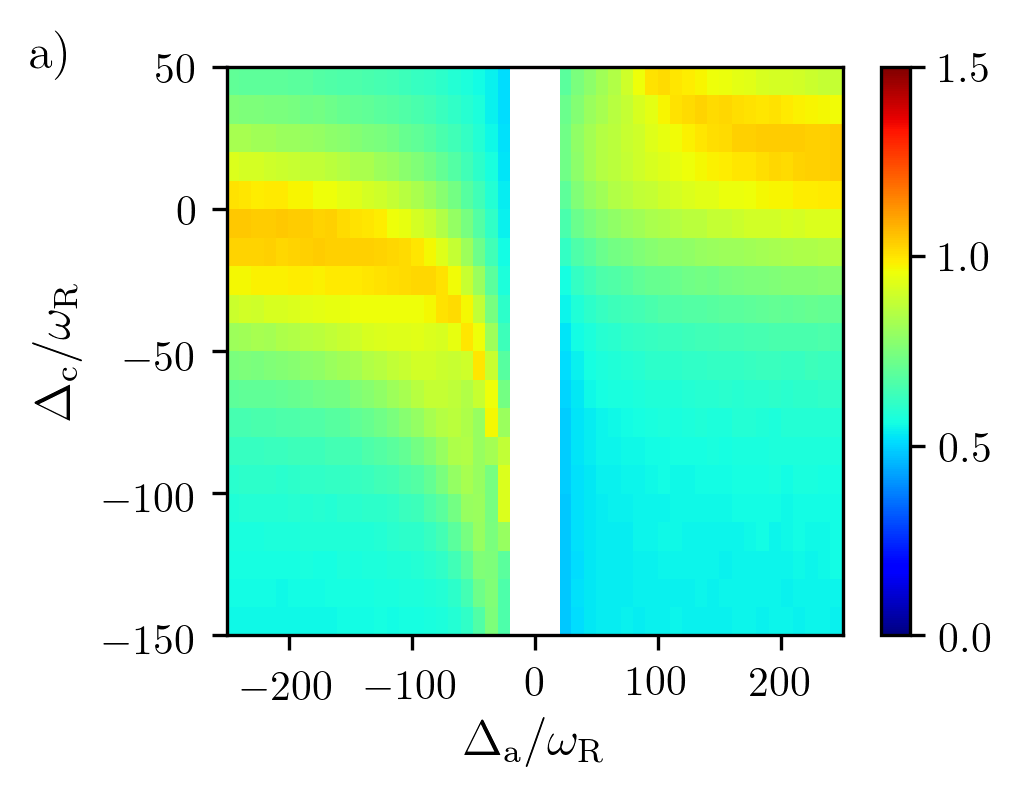}
\includegraphics[width=.32\linewidth]{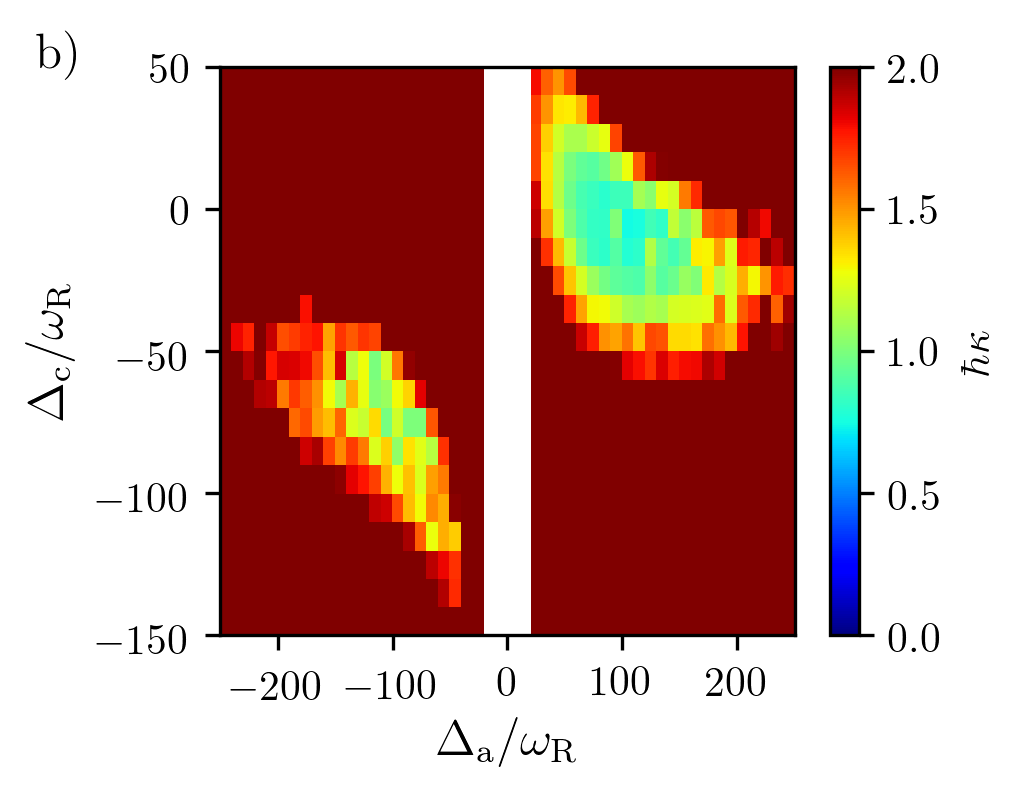}
\includegraphics[width=.32\linewidth]{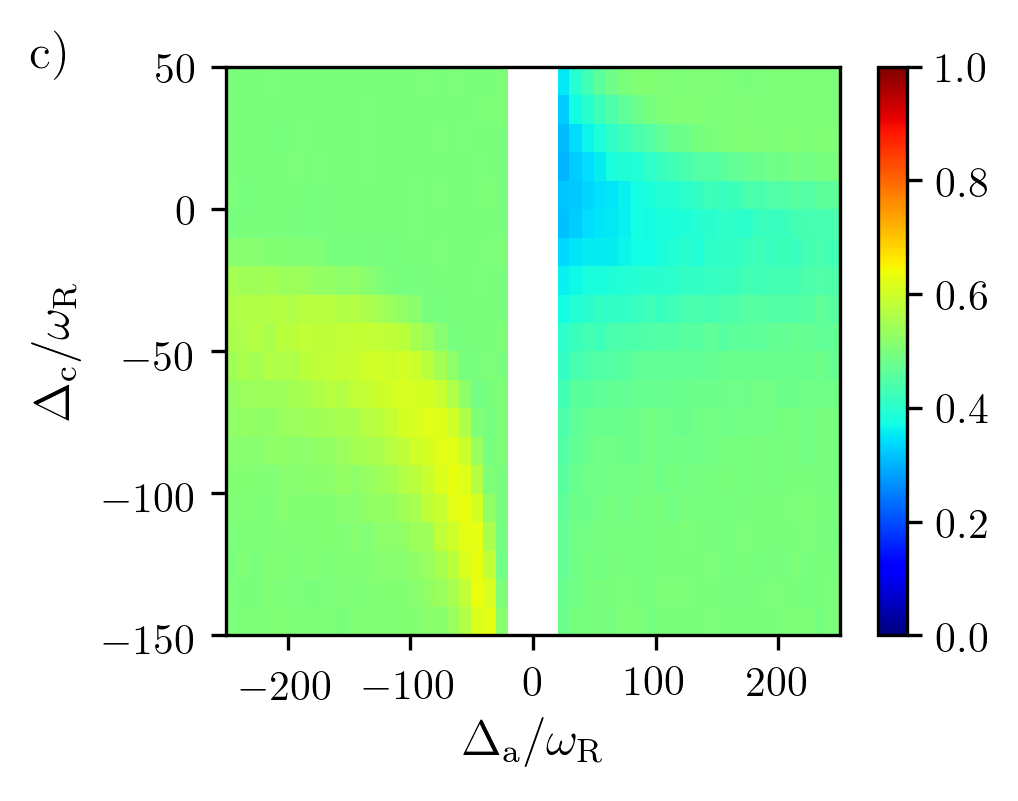}
\caption{\small $\Delta_{\mathrm{a}}$, $\Delta_{\mathrm{c}}$ scan of a) the mean intensity $\Intensity$, b) the mean kinetic energy $\Ekin/\hbar\kappa$ and c) the bunching parameter $\bunch$ for a single particle in a longitudinally pumped cavity, with $\eta_{\mathrm{l}}=30\,\omega_{\mathrm{R}}$. Energies higher than $2\hbar\kappa$ were cut away.}
\label{Fig.LongDaDc1}
\end{figure*}

The scan of the intra-cavity steady state intensity $\left<\left|\alpha\right| ^2 \right>$ (Fig.$\,$\ref{Fig.LongDaDc1}a) indicates the amount of light interacting with the atom on average. From Eq.~\eqref{Eq.Alpha} we can estimate the steady state intensity as  
\begin{align}
    \left|\alpha_{\mathrm{st}} \right|^2 = \frac{\eta_\mathrm{l}^2}{\kappa^2+\Delta_{\mathrm{eff}}^2},
    \label{Eq.SteadyInt}
\end{align}
with $\Delta_{\mathrm{eff}}=\Delta_{c}-U_{0}\cos^2\left(kx\right)$. For the majority of the here chosen parameters the intensity is $\leq 1$. Although the intensity is so low on average, the kinetic energy scan shown in Fig.$\,$\ref{Fig.LongDaDc1}\,b) reveals tuples of detunig parameters $\left(\Delta_{\mathrm{a}},\Delta_{\mathrm{c}}\right)$ for which the particle has been slowed down from initially $\Ekin = 7.5\,\hbar\kappa$ to $\hbar\kappa$. Moreover, the cooling region is broader for $\Delta_{\mathrm{a}}>0$, which arises from the dispersive red-shift of the effective cavity detuning, and thus allows to transfer a higher amount of the particle kinetic energy to the cavity field.
Fig.~\ref{Fig.LongDaDc1}\,c) shows the localization position of the particle, which is estimated by the bunching parameter $\bunch=\left<\cos^2\left(kx \right)\right>$. Blue regions indicate trapping at the intensity minima of the cavity field, as $\bunch \rightarrow 0$ and red regions indicate trapping at the intensity maxima, as $\bunch \rightarrow 1$. The confinement for the here chosen parameters is rather weak, for red as well as blue detuning in the areas of minimal $\Ekin$. For blue detuned pump we obtain $\bunch \approx 0.37$, for red detuned pump $\bunch \approx 0.59$, at best. A value of $\bunch=0.5$ indicates no trapping. Thus, the here obtained values indicate a very weak confinement, which can be improved by increasing the depth of the cavity potential. Therefore, the in the next section we will analyze the dependence of the system on the pump strength. Choosing $\Delta_{\mathrm{c}}=-40\,\omega_{\mathrm{R}}$ turned out to be suitable for comparison between red and blue detuned pump.

\subsubsection{\label{sec:longDaEta}Atomic detuning vs.\ pump strength - scan}

\begin{figure*}[t]
\centering
\includegraphics[width=.32\linewidth]{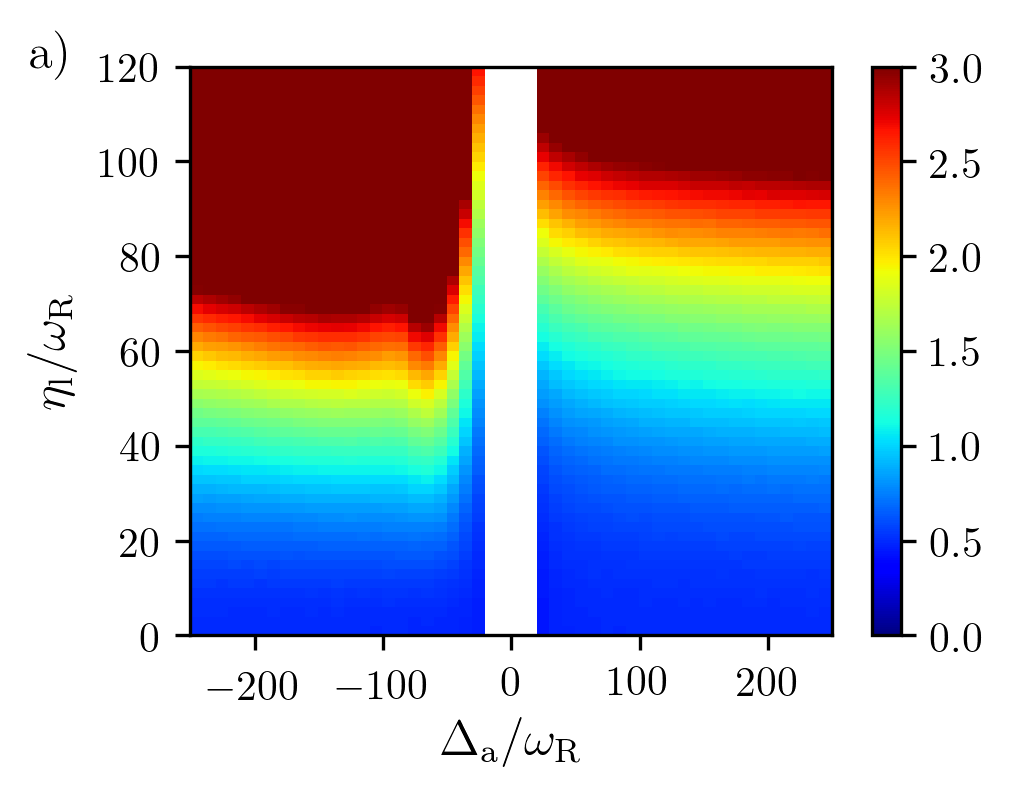}
\includegraphics[width=.32\linewidth]{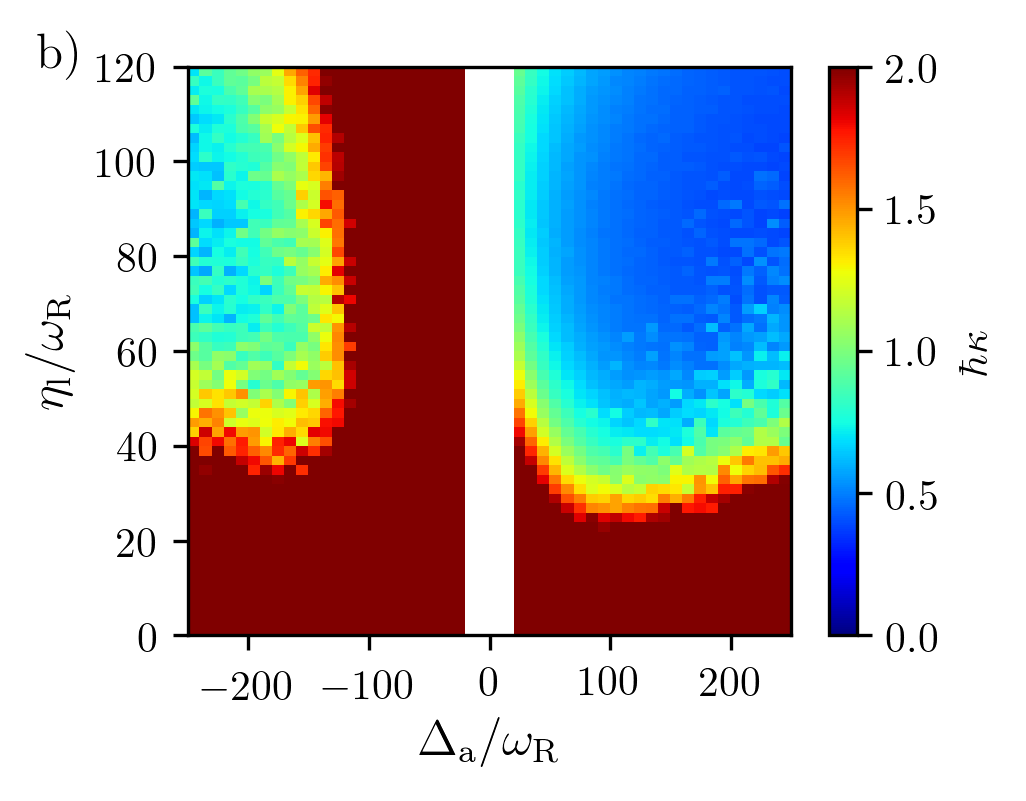}
\includegraphics[width=.32\linewidth]{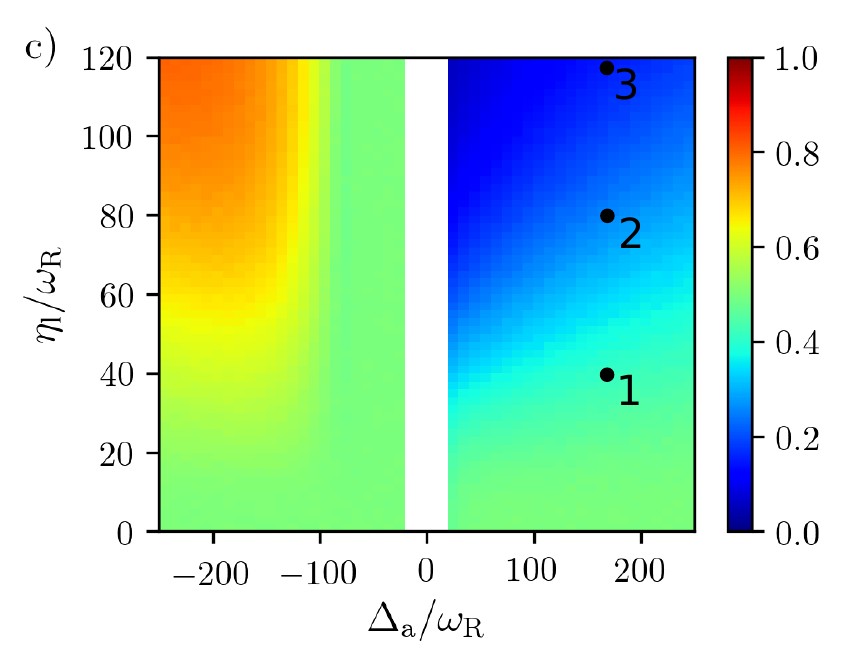}
\caption{$\Delta_{\mathrm{a}}$, $\eta_{\mathrm{l}}$ scan of a) the mean intensity $\Intensity$, b) the mean kinetic energy $\Ekin$ and c) the bunching parameter $\bunch$ for a single particle in a longitudinally pumped linear cavity, with $\Delta_{\mathrm{c}}= -\kappa$. Energies higher than $2\hbar\kappa$ were cut away.}
\label{Fig.LongDaEta1}
\end{figure*}

The cavity intensity, shown in Fig.\,\ref{Fig.LongDaEta1}\,a) is monotonically increasing with the pump strength according to Eq.\,\eqref{Eq.SteadyInt} for both red and blue detuning. For red detuning the atom arranges in a way that increases the mode intensity, whereas for blue detuning it arranges such that $\Delta_{\mathrm{eff}} \rightarrow \Delta_{\mathrm{c}}$, as $\cos^2\left(kx\right) \rightarrow 0$, see Fig.\,\ref{Fig.LongDaEta1}\,c). Thus, the intensity for red detuned pump for any $\eta_{\mathrm{l}}\neq 0$ equals twice the intensity of a blue detuned pump at the same pump strength. Moreover, a blue detuned pump allows to reach kinetic energies as low as $\Ekin=\hbar\kappa/2$, see Fig.$\,$\ref{Fig.LongDaEta1}\,b), over a very broad interval of $\eta_\mathrm{l}$ and $\Delta_{\mathrm{a}}$ values. In this case the kinetic temperature reaches the value $k_{\mathrm{B}}T=\hbar\kappa$. For red detuned pump light the minimal steady state kinetic energy for the here analyzed parameters is approximately a factor of two higher than for blue detuning. 
Fig.$\,$\ref{Fig.LongDaEta1}\,c) shows organization of the particle in the high intensity regions for red detuned pump, as expected. Organization occurs also for blue detuned light, where the particle tend towards the intensity minima of the cavity mode, indicated by the blue coloration. Similar to red detuning, also here organization sets in after a certain cavity intensity is reached. From the picture we deduce that an increasing pump rate leads to a better confinement of the particle. 

\subsubsection{\label{sec:longTrajDist}Trajectories and position distribution}
\begin{figure}[t]
\centering
\includegraphics[width=.8\linewidth]{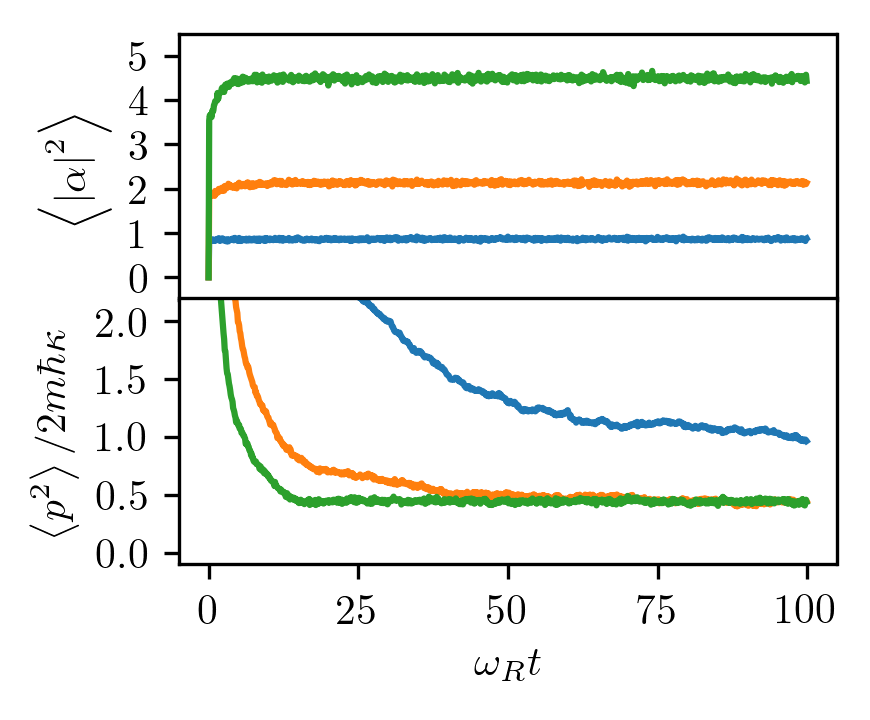}
\caption{Averaged trajectories of the intra-cavity intensity (top) and particle kinetic energy (bottom) for the parameters $\Delta_{\mathrm{a}}=180\omega_{\mathrm{R}}$, blue: $\eta_{\mathrm{l}}=40\,\omega_{\mathrm{R}}$, orange: $\eta_{\mathrm{l}}=80\,\omega_{\mathrm{R}}$ and green: $\eta_{\mathrm{l}}=120\,\omega_{\mathrm{R}}$ (see Fig.$\,$\ref{Fig.LongDaEta1}). The average is taken over $2000$ trajectories, where for each trajectory the initial position of the particle in the cavity potential and its direction of motion are chosen randomly.}
\label{Fig.LongDaEtaTrajectories1}
\end{figure}

\begin{figure}[t]
\centering
\includegraphics[width=1.\linewidth]{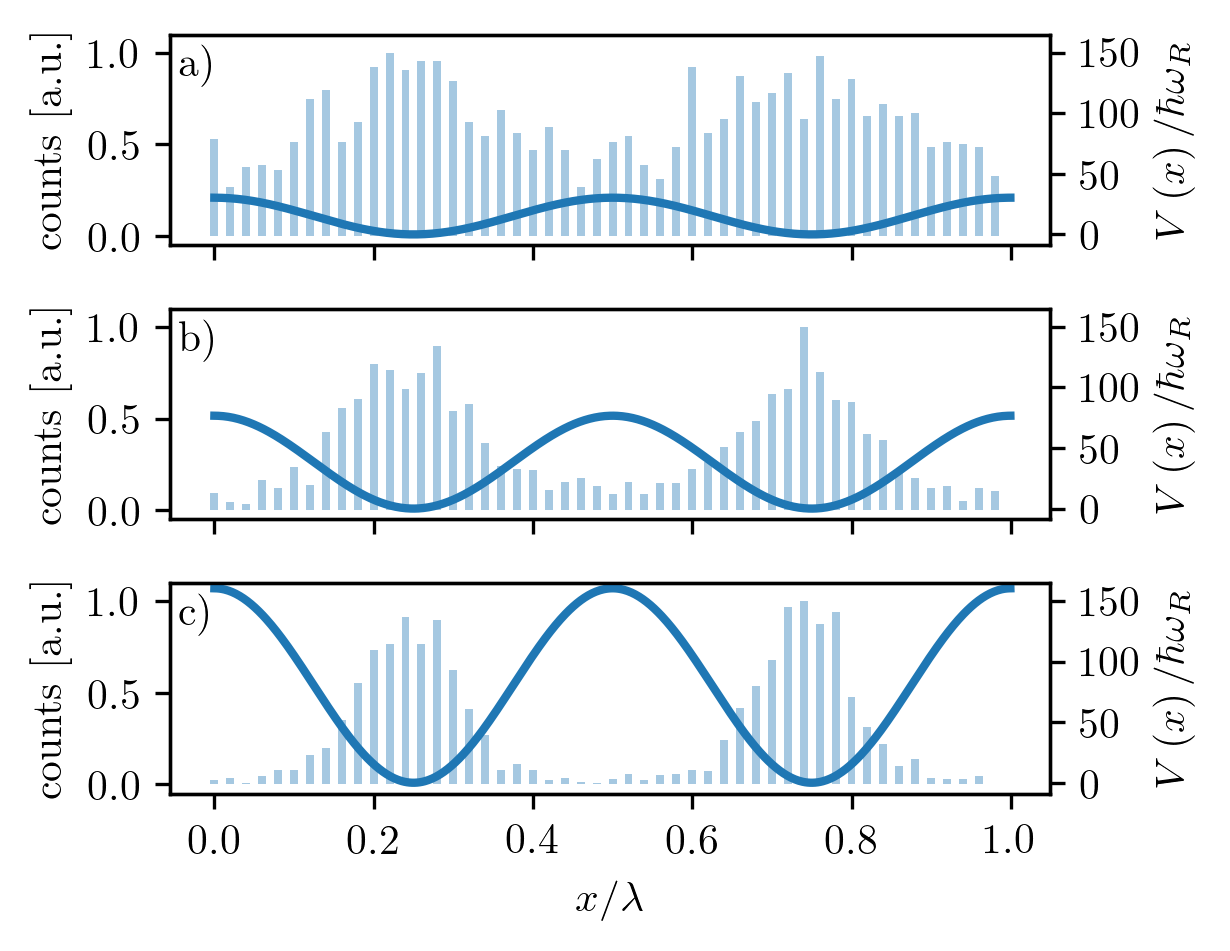}
\caption{Steady state single particle position distribution (horizontal bars) in the cavity potential \eqref{Eq.Potential} of a longitudinally pumped linear cavity (solid blue line) for $\Delta_{\mathrm{a}}=180\,\omega_{\mathrm{R}}$, $\eta_{\mathrm{l}}=40\,\omega_{\mathrm{R}}$ a), $\eta_{\mathrm{l}}=80\,\omega_{\mathrm{R}}$ b) and $\eta_{\mathrm{l}}=120\,\omega_{\mathrm{R}}$ c). The statistics show the position of the particle at $100\,\omega_{\mathrm{R}}^{-1}$ over $10000$ trajectories. For each trajectory the initial position of the particle and its direction of motion were chosen randomly.}
\label{Fig.LongDaEtaStatistics1}
\end{figure}

In the following we analyze the effect of the intra-cavity intensity onto the cooling and trapping properties by the example of three points indicated in the blue detuned region of Fig.$\,$\ref{Fig.LongDaEta1}\,c). 

The averaged temporal evolution of cavity mode intensity and the particle energy for the three pump values are shown in the top and bottom panel of Fig.$\,$\ref{Fig.LongDaEtaTrajectories1}. The cavity mode reaches a steady after very few $\omega_{\mathrm{R}}^{-1}$ for each point and remains constant for later times. Thus, it is possible to estimate an average optical potential depth using the steady state intensity value in Eq.\,\eqref{Eq.Potential}. The cooling however, depends strongly on the amount of light in the cavity. Although, each of the here shown trajectories reaches a kinetic energy $\leq \hbar\kappa$ (see Fig.$\,$\ref{Fig.LongDaEtaTrajectories1} bottom panel), it takes significantly longer to cool the particle, if the cavity intensity is low. As we can see, for the lowest pump strength, the particle could not be cooled to the steady state in the time interval shown in the figure, whereas higher pump-rates allow cooling to kinetic energies as low as $\hbar\kappa/2$ within less than $40\,\omega_{\mathrm{R}}^{-1}$.

In Fig.$\,$\ref{Fig.LongDaEtaStatistics1} we show the position distribution of the particle inside the optical potential (solid line) for three different pump-rates. In the case of blue detuning, a modulation on the particle distribution can be observed for a very low pump-rate. However, since the particle's kinetic energy did not reach the steady state for this parameters, we see rather an increased probability of finding the particle at the nodes of the cavity field due to the reduction of the velocity while passing through the potential, than a strong confinement inside the potential well (Fig.$\,$\ref{Fig.LongDaEtaStatistics1}\,a). An increasing pump strength leads to a higher cavity mode intensity and therefore also the potential depth increases on average. As the steady state kinetic energy of the particle becomes small compared to the potential depth, the probability of finding the particle at positions other than the nodes of the modes is highly reduced, as well as the probability of the particle jumping over the potential wall. Thus, Figs.$\,$\ref{Fig.LongDaEtaStatistics1}\,b) and c) show the evolution of the particle localization towards a stronger confinement at the field nodes with increasing pump strength.

\subsection{\label{sec:Trans}Transverse pumping of the atom}

In the following section we discuss the behaviour of the system, for $\eta_{\mathrm{l}}=0$ and $\eta_{\mathrm{l}}>0$. For the simulations we fixed the position of the particle at the anti-node of the pump field, which can be achieved with an additional far off-resonant trap in an experiment.

\subsubsection{\label{sec:transDaDc}Atom detuning vs.\ cavity detuning - scan}

\begin{figure*}[t]
\centering
\includegraphics[width=.32\linewidth]{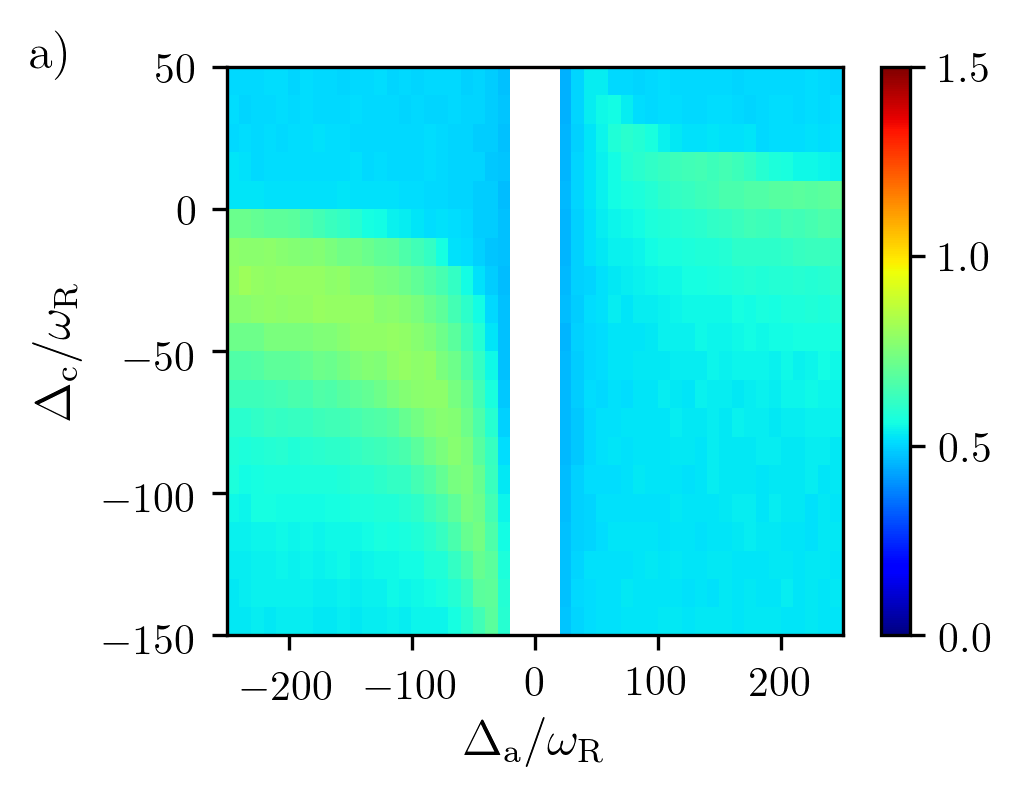}
\includegraphics[width=.32\linewidth]{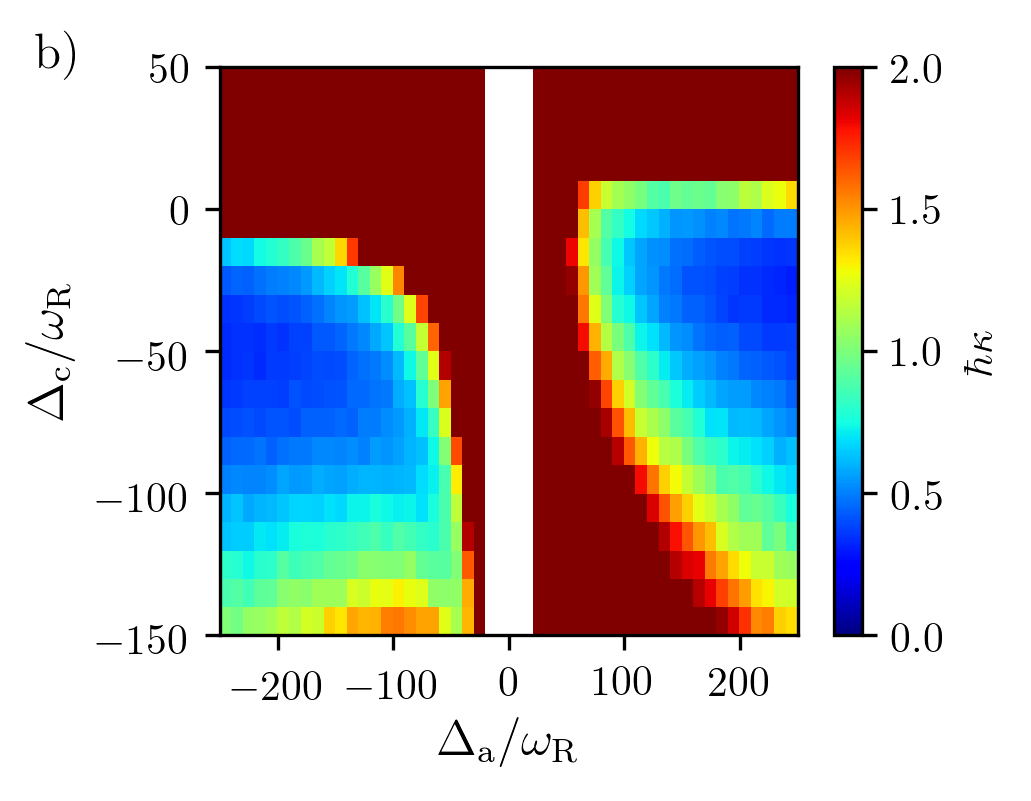}
\includegraphics[width=.32\linewidth]{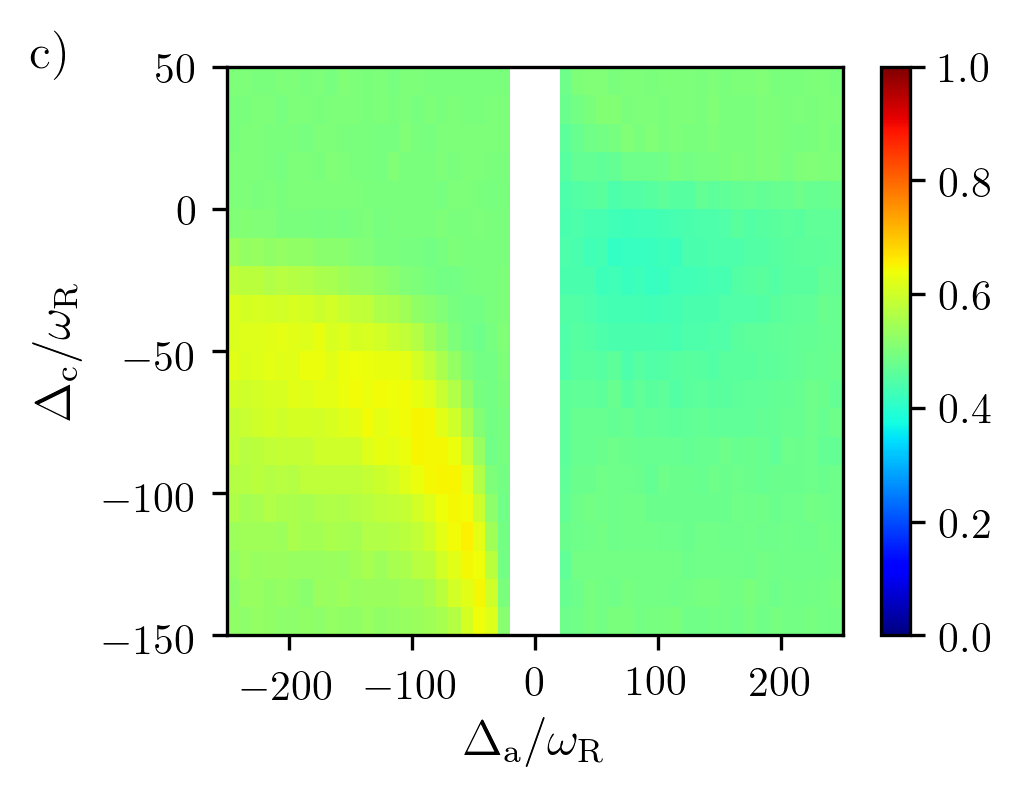}
\caption{$\Delta_{\mathrm{a}}$, $\Delta_{\mathrm{c}}$ scan of a) the mean intensity $\Intensity$, b) the mean kinetic energy $\Ekin$ and c) the bunching parameter $\bunch$ for a single particle in a transversely pumped cavity, with $\eta_{\mathrm{eff}}=30\,\omega_{\mathrm{R}}$. Energies higher than $2\hbar\kappa$ were cut away.}
\label{Fig.TransDaDc1}
\end{figure*}

The intensity scan in Fig.\,\ref{Fig.TransDaDc1}\,a) shows a maximal steady state intensity of $\Intensity \approx 0.8$, which is predominantly found around the cooling and trapping region for red detuning. It is well known, that in this regime the atoms arrange in a way that enhances light scattering into the cavity mode. For $\Delta_{\mathrm{a}} > 0$ the intra-cavity intensity in the cooling and trapping regions is approximately a factor of two lower than for red detuning, indicating an arrangement of the particle in a way that minimizes the intra-cavity intensity. Moreover, compared to the cavity pump geometry treated in Sec.\,\ref{sec:Long}, a transverse pump allows to cool the particle for both detunigs down to $\Ekin \approx \hbar\kappa/4$, which corresponds to a kinetic temperature of $k_{\mathrm{B}}T = \hbar\kappa/2$, see Fig.\,\ref{Fig.TransDaDc1}\,b). Furthermore, cooling is achieved for a much broader range of detuning parameters for both regimes. Although the low cavity intensity suggests a good confinement of the particle for blue detuned light, Fig.\,\ref{Fig.TransDaDc1}\,c) shows only a confinement of the particle for red detuned pump at first sight. A very careful look onto the bunching parameter scan allows to recognize a narrow light blue region on the right hand side of the figure. The bunching parameter reaches the values $\bunch \approx 0.65$ for red and $\bunch = 0.41$ for blue detuned pump. These values are similar to the ones obtained in the case of a longitudinal pump. 

For both geometries a pump strength of $30\,\omega_{\mathrm{R}}$ is sufficient to cause a strong reduction of the particle's kinetic energy. Moreover, tuples of cavity and atomic-detuning parameters were found, where even a low pump strength leads to a confinement on the particle. Also for the case of a transverse pump $\Delta_{\mathrm{c}} = - \kappa$ turned out to be a good candidate for finding both, cooling and trapping.

\subsubsection{\label{sec:transDaEta}Atomic detuning vs.\ pump strength - scan}

\begin{figure*}[t]
\centering 
\includegraphics[width=.32\linewidth]{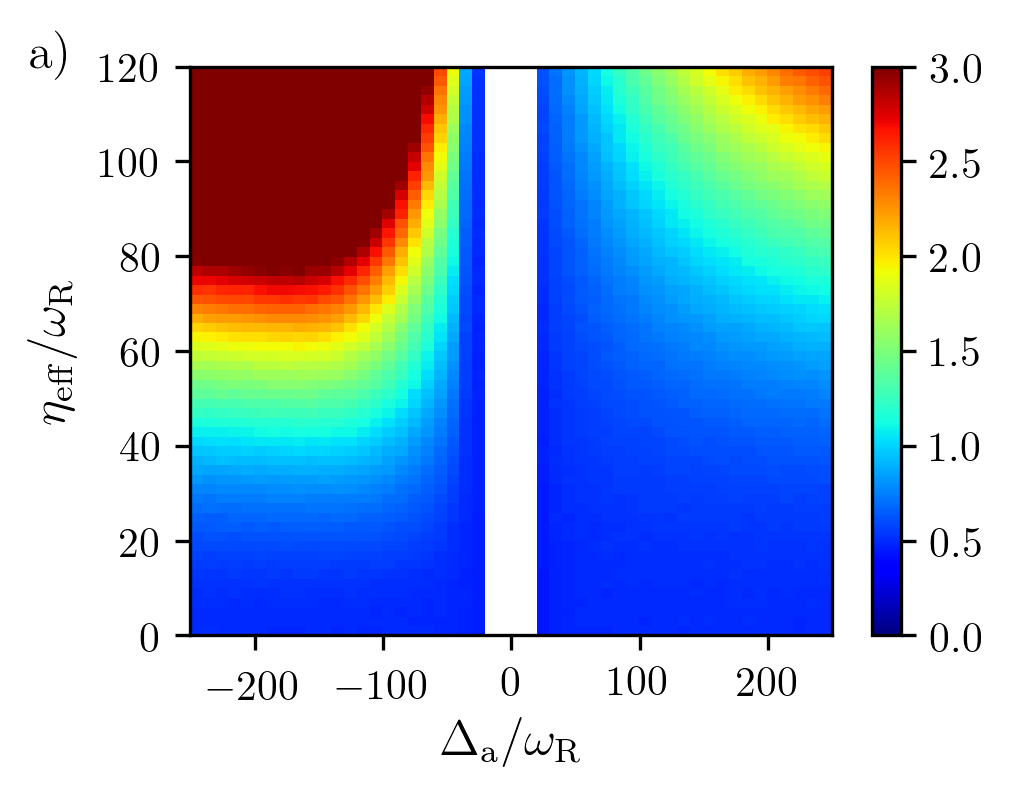}
\includegraphics[width=.32\linewidth]{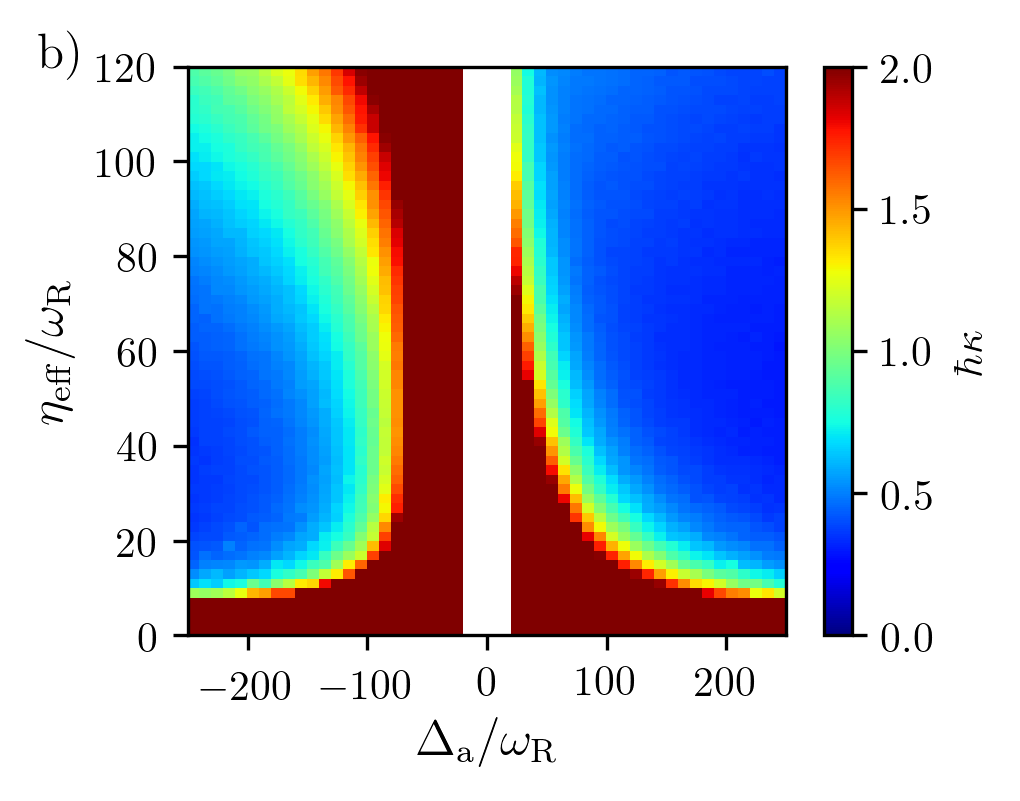}
\includegraphics[width=.32\linewidth]{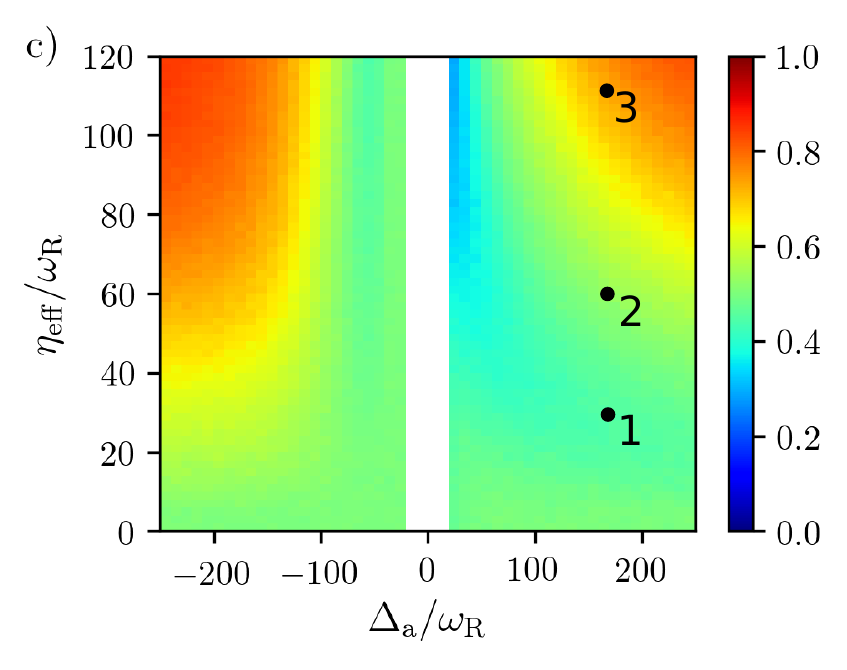}
\caption{$\Delta_{\mathrm{a}}$, $\eta_{\mathrm{eff}}$ scan of a) the mean intensity $\Intensity$, b) the mean kinetic energy $\Ekin$ and c) the bunching parameter $\bunch$ for a single particle in a transverse pump setup, with $\Delta_{\mathrm{c}}= -\kappa$. Energies higher than $2\hbar\kappa$ were cut away.}
\label{Fig.TransDaEta1}
\end{figure*}

Fig.\,\ref{Fig.TransDaEta1}\,c) shows the same threshold behaviour of the bunching parameter for red detuning, as it was described in \cite{niedenzu2011kinetic}. Surpassing a critical pump strength leads to a self ordering of the particle, as the bunching parameter jumps from the value $0.5$ to $1$. The particle localizes at the intensity maxima of the cavity field and enhances light scattering into the cavity, see Fig.~\ref{Fig.TransDaEta1}\,a). For red detuned pump the particle is cooled to a lower steady state kinetic energy before the self-organization occurs Fig.~\ref{Fig.TransDaEta1}\,b). The kinetic energy of the trapped particle is increased, as explained in \cite{niedenzu2011kinetic}. 

For small, blue atomic detunings of $\Delta_{\mathrm{a}} < 100\,\omega_{\mathrm{R}}$ we observe self-trapping of the particle at the nodes of the cavity field, as it can be seen from the bunching parameter in Fig.\,\ref{Fig.TransDaEta1}\,c). Also the low mode-intensity shown in Fig.\,\ref{Fig.TransDaEta1}\,a) for $\kappa < \Delta_{\mathrm{a}} < 100\,\omega_{\mathrm{R}}$ enforces this assumption, since for $\bunch \rightarrow 0$ the intra-cavity steady state intensity tends towards zero, too. As the pump rate and the atomic detuning increase, for blue detuned pump light, the bunching parameter starts increasing. For $\Delta_{\mathrm{a}} > 100\,\omega_{\mathrm{R}}$ and $\eta_{\mathrm{eff}} > 80\,\omega_{\mathrm{R}}$ the bunching parameter makes a transition towards $\bunch \rightarrow 1$. This indicates a reordering of the particle to the anti-node of the cavity field. In this case also the amount of light scattered into the cavity mode increases. The kinetic energy for blue detuned pump remains around $\hbar\kappa/4$ for the most of the here chosen parameters, in contrast to red detuning.

\subsubsection{\label{sec:transTrajDist}Trajectories and position distribution}

\begin{figure}[t]
\centering
\includegraphics[width=.8\linewidth]{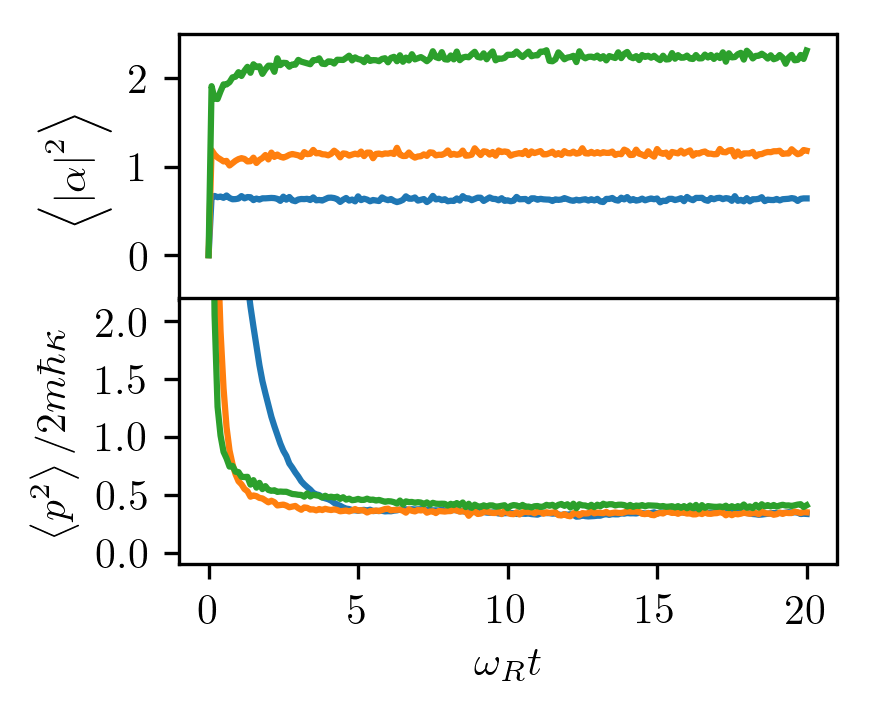}
\caption{Similar to Fig.\,\ref{Fig.LongDaEtaTrajectories1}, with blue: $\eta_{\mathrm{eff}}=30\,\omega_{\mathrm{R}}$, orange: $\eta_{\mathrm{eff}}=60\,\omega_{\mathrm{R}}$ and green: $\eta_{\mathrm{eff}}=110\,\omega_{\mathrm{R}}$.}
\label{Fig.TransDaEtaTrajectories1}
\end{figure}

Similar to the previous section Fig.$\,$\ref{Fig.TransDaEtaTrajectories1} shows the temporal evolution of the averaged intensity as well as the averaged kinetic energy, respectively. For a weak pump the cavity intensity is below one, as $\eta_{\mathrm{eff}}< \kappa$ but also the particle tends towards the nodes of the cavity mode where light scattering into the mode is suppressed. As we have seen in the previous section, with increasing pump strength the particle is trapped at the anti-nodes of the cavity mode, enhancing the cavity intensity. The transition of the positioning of the particle is caused by the change of the shape of the optical potential. 

\begin{figure}[t]
\centering
\includegraphics[width=1.\linewidth]{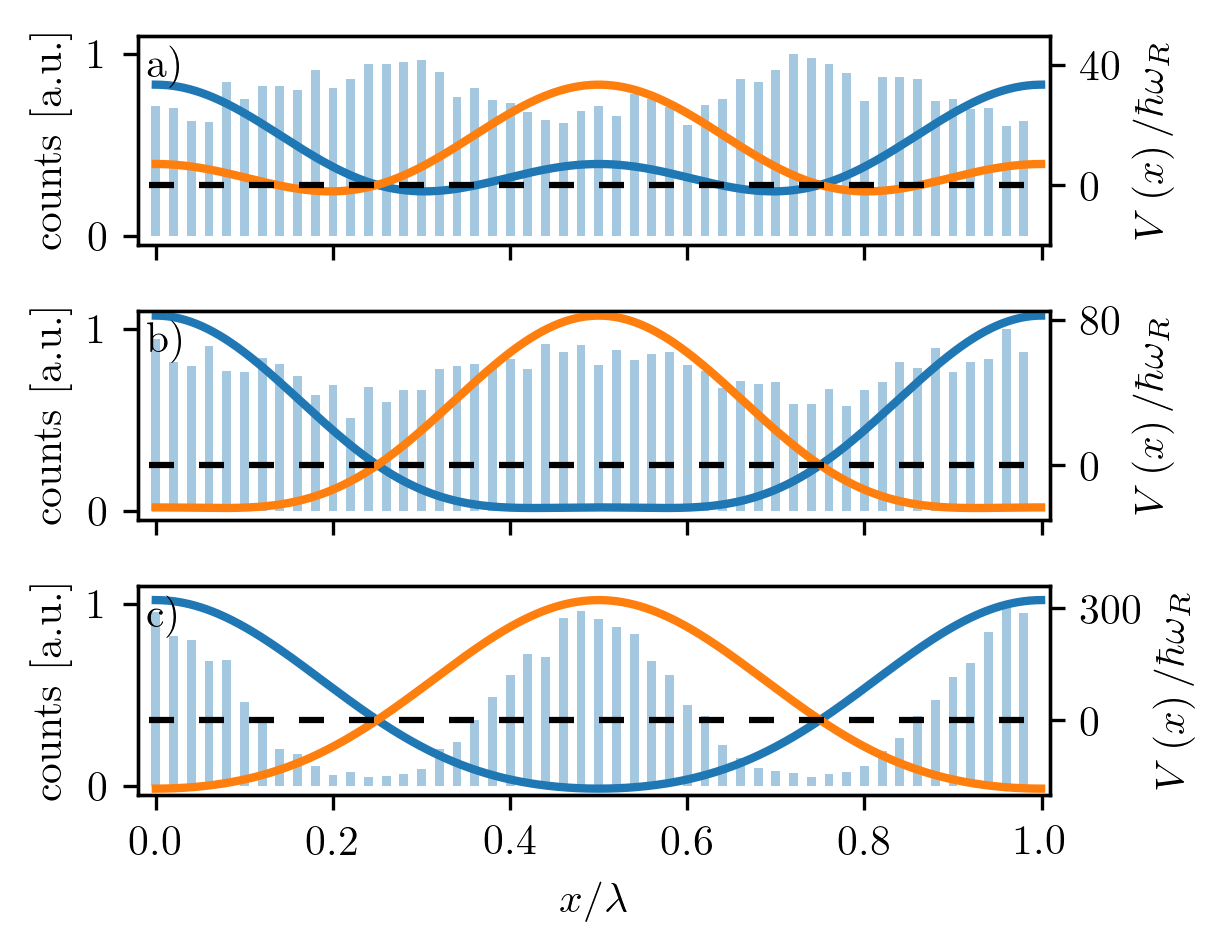}
\caption{Similar to Fig.\,\ref{Fig.LongDaEtaStatistics1}, with 1) $\eta_{\mathrm{t}}=30\,\omega_{\mathrm{R}}$, 2) $\eta_{\mathrm{t}}=60\,\omega_{\mathrm{R}}$ and 3) $\eta_{\mathrm{t}}=110\,\omega_{\mathrm{R}}$. The blue solid line indicates a potential with $\realp\left(\alpha\right)>0$, the orange solid line with $\realp\left(\alpha\right)<0$. Note the increasing potential depth in the sub-figures.}
\label{Fig.TransDaEtaStatistics1}
\end{figure}

Fig.\,\ref{Fig.TransDaEtaStatistics1} shows the position distribution of the particle in the potential created by scattering light off the atom into the cavity. The optical potential Eq.\,\eqref{Eq.Potential} consist of two contributions, one proportional to $\cos^2\left( kx \right)$, arising from the cavity field and one proportional to $\cos(kx)$, arising from the interference between cavity and pump field \cite{ritsch2013cold}. The phase of the latter expression is highly dependent on the position of the atom and the phase of the scattered light. The resulting potential shape is indicated by the blue and orange solid line in the figure for $\realp\left( \alpha \right)$ positive and negative, respectively. In the case of a weak pump the cavity potential dominates and the action of the potential is mostly repulsive. In this case the particle is pushed towards the local potential minima at approximately $x=(2n-1)\pi/4$, with $n \in \mathbb{N}$, where the cavity mode intensity is minimal. As however, the $\eta_{\mathrm{eff}}$ increases, the interference term gains dominance and the potential takes a $\cos(kx)$ shape, with attractive contributions. The particle is then drawn to $x=n\lambda/2$, which coincide with position of maximal mode intensity.

The total intensity inside the cavity is~\cite{asboth2005self}
\begin{equation}
    I_{\mathrm{tot}} \propto \left|\alpha \cos(kx) + \Omega/g_0 \right|^2.
\end{equation}
 Thus, although on first sight the position of the particle coincides now with the intensity maximum of the cavity mode, the total intensity inside the cavity is minimal, since the particle arranges such that the interference term between cavity field and pump is destructive (see Fig.\,\ref{Fig.TransDaEtaStatistics1}). As a consequence the total intensity is reduced and the particle remains low field seeking.

The localization of the particle could not be fixed to either of the two possible self-induced potentials. We observed a jumping of the particle between the minima of the blue and orange potential. These arise from the fluctuations on the cavity field as well as the particle motion, because the phase of the scattered light highly depends those. Moreover, minimal fluctuations at these intensity values can lead to a complete annihilation of the cavity field and thus, destruction of the confinement on the particle.

\section{\label{sec:Coolingtime} Cooling time}
\begin{figure}[t]
\centering
\includegraphics[width=.48\linewidth]{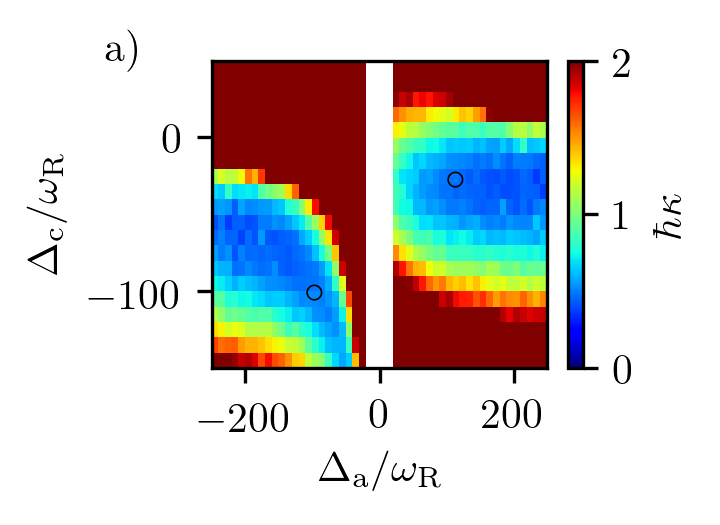}
\includegraphics[width=.48\linewidth]{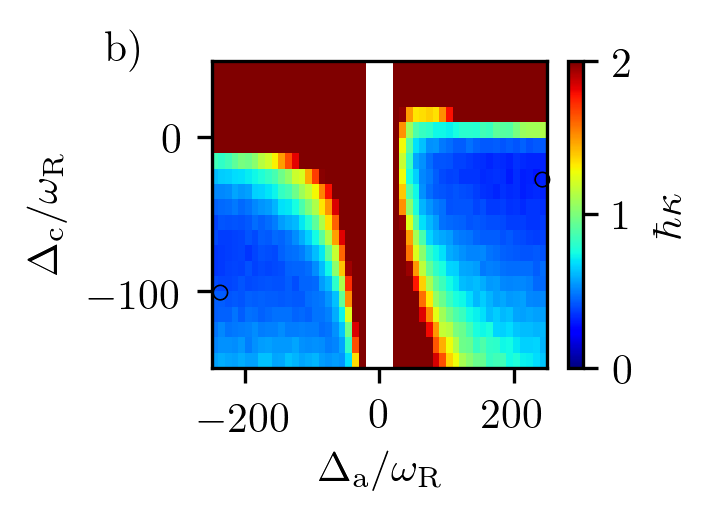}\\
\includegraphics[width=.48\linewidth]{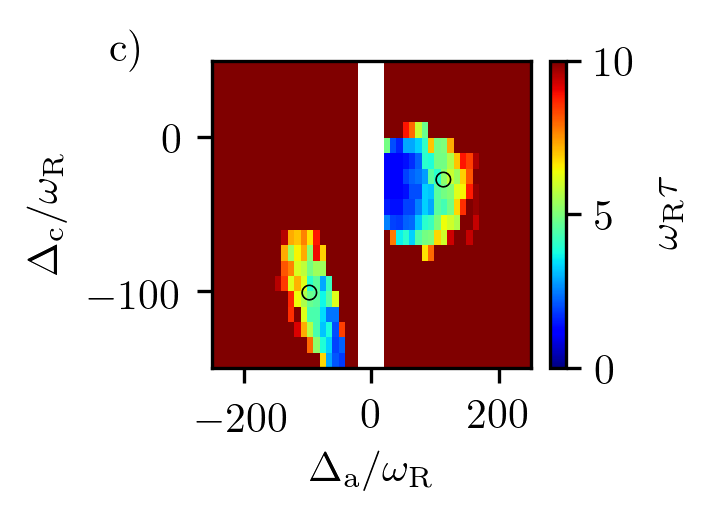}
\includegraphics[width=.48\linewidth]{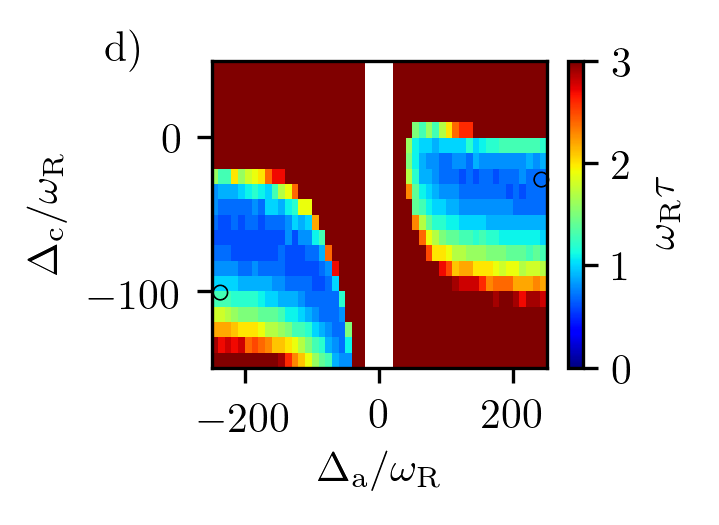}
\caption{Kinetic energy for cavity a) and atom pump b), and cooling time for cavity c) and atom pump d). Here, we chose $\eta_{\mathrm{l}} = 80\omega_{\mathrm{R}} = \eta_{\mathrm{eff}}$. Note the difference in cooling time between cavity and atom pump geometry and the different color coding scales in the two lower pictures.}
\label{Fig.CoolingTime}
\end{figure}

We have shown for which detuning and pump parameters we expect cavity cooling and trapping. However, it is also of interest to know how fast kinetic energy can be reduced. Therefore, in Fig.\,\ref{Fig.CoolingTime} we show a scan of the kinetic energy (top) vs. the time (bottom) it takes to reduce kinetic energy to the value $\hbar\kappa$ for a cavity pump (a and c) and an atom pump (b and d) geometry. We only consider parameters, for which the steady state kinetic energy is below $\hbar\kappa$. In particular we want to compare the pump geometries and the respective detuning regimes. 

The cooling time shows a strong dependence on the detuning parameters $\Delta_{\mathrm{a}}$ and $\Delta_{\mathrm{c}}$, as they change the effective detuning $\Delta_{\mathrm{eff}}$ Eq.\,\eqref{Eq.SteadyInt}. Dependent on the sign of $\Delta_{\mathrm{eff}}$ the cavity has either a heating or a cooling effect. A larger red detuning of the pump with respect to the cavity allows to transfer a higher amount of kinetic energy from the atom to the cavity field. However, the dispersive shift should not exceed the cavity linewidth, since then energy cannot be dissipated through the cavity emission channel and leads to a heating of the particle. This trade-off therefore leads to a competition between low steady state energy, low saturation and fast cooling. We have seen for red detuning the region of minimal cooling time coincides with high atomic saturation and is therefore inconsiderable, especially for the longitudinal case. For blue detuning however, low energy regions overlap very well with low saturation and fast cooling regions.

\begin{figure}[t]
\centering
\includegraphics[width=.8\linewidth]{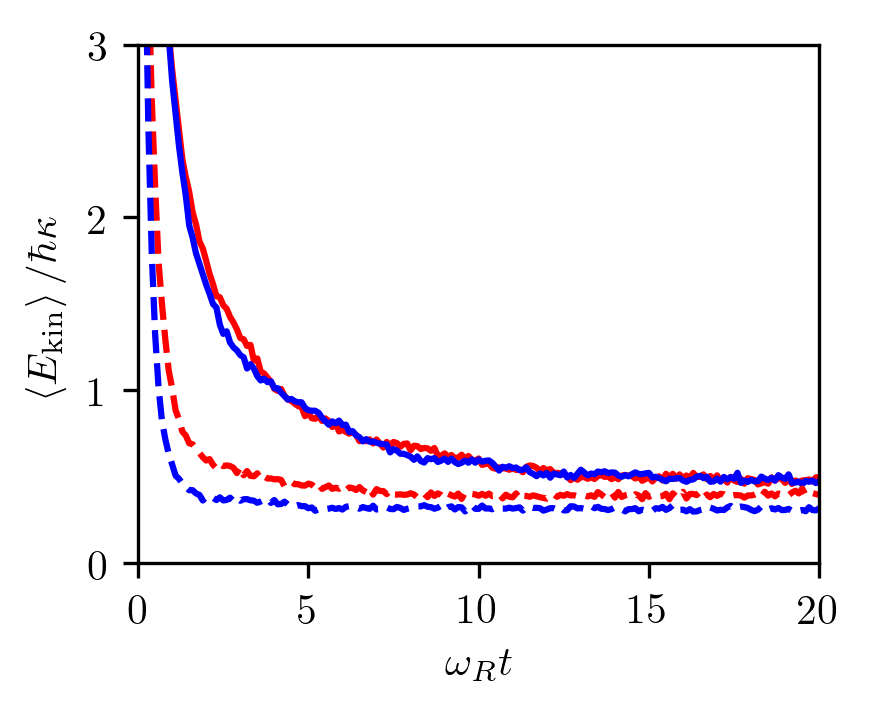}
\caption{Average kinetic energy evolution for longitudinal cavity pump (solid line) with $\left|\Delta_{\mathrm{a}}\right|=100\,\omega_{\mathrm{R}}$ and transverse atom pump (dashed line) with $\left|\Delta_{\mathrm{a}}\right|=250\,\omega_{\mathrm{R}}$. The red curve represents red detuning with $\Delta_{\mathrm{c}}^{\mathrm{(red)}}=-100\,\omega_{\mathrm{R}}$ and blue curve represents blue detuning with $\Delta_{\mathrm{c}}^{\mathrm{(blue)}}=-30\,\omega_{\mathrm{R}}$.}
\label{Fig.LongCoolingTimeTrajectory}
\end{figure}

In Fig.\,\ref{Fig.LongCoolingTimeTrajectory} the time evolution of the averaged kinetic energy is shown for the parameters indicated by circles in Fig.\,\ref{Fig.CoolingTime} for the respective geometries and detunings. In both cases parameters were chosen where the low-saturation condition $s \left|\alpha\right|^2 \ll 1$ is fulfilled and the steady state kinetic energy reaches values $\Ekin \leq \hbar\kappa/2$. According to~\cite{niedenzu2011kinetic} for transverse pump the kinetic temperature is estimated by
\begin{align}
k_{\mathrm{B}}T=\hbar\frac{\kappa^2+\Delta_{\mathrm{eff}}^2}{4\left|\Delta_{\mathrm{eff}}\right|,}
\end{align}
for $\Gamma \ll \kappa$. The here found steady state values coincide with the values obtained by above formula for both pump geometries. The results show that most of the kinetic energy is lost within $10\,\omega_{\mathrm{R}}^{-1}$. For both detunings in either geometry parameters can be found for which the cooling time is minimal. Comparing the average time it takes to cool for these parameters, we see that there is barely a difference between red and blue detuning, as for the choice of $\Delta_{\mathrm{eff}}$. A higher influence on cooling time is given by the pump geometry. Whereas for a longitudinal cavity pump it takes several $\omega_{\mathrm{R}}^{-1}$ to get rid of a major part of the kinetic energy, a transverse atom pump can be up to $10$ times faster. For the here chosen parameters the difference is a factor of $5$. The origin of this discrepancy might be connected to the total depth of the optical potential, as a comparison between Figs.\,\ref{Fig.LongDaEtaStatistics1} and \ref{Fig.TransDaEtaStatistics1} indicates a much deeper potential when the system is pumped from the side.

\section{\label{sec:Conclusions} Conclusions}
We revisited the field of point particle cavity cooling with new emphasis on the case of low field seeking particles, which eventually can be cooled and trapped near nodes of the field with minimum perturbation and intrinsic heating. Using semi-classical point particle simulations including spontaneous emission and cavity loss we identify the parameter regimes where fast cooling concurs with trapping and low kinetic temperatures in the vicinity of field minima or even zeros.

As a central result we find kinetic temperatures even lower than for red detuning at much lower saturation and even faster cooling rates. Despite the fact that the particles sit close to zero field, self-trapping via transverse pump scattering to a cavity field can be generalized to blue detuning as well. While the particles for weak pump still sit at the cavity field nodes, higher pump powers exhibit a transition to trapping at the antinodes, where pump and cavity field destructively interfere to create a local minimum. A closely related behaviour was recently predicted as well for trapping a BEC at zero temperature in a cavity field \cite{piazza2015self}.

Besides faster cooling and lower temperatures, trapping at zero field can also provide advantages for spectroscopic applications or cooling of molecules which are sensitive to strong fields. While this should be also true for nano-particles it is not so obvious how a low field seeking interaction can be implemented in this case. 
So far we mainly studied the single particle limit, but of course the scaling properties of blue cavity cooling with particle number need further consideration as well to turn it into a practical procedure. At the end using two very different cavity modes blue and red cooling might be combined to achieve fast cooling and deep trapping simultaneously.   

\section*{Acknowledgements}
This work was performed in the framework of the European Training Network ColOpt, which is funded by the European Union (EU) Horizon 2020 programme under the Marie Skłodowska-Curie action, grant agreement 721465. We acknowledge support by the Austrian Science Fund FWF within the DK-ALM (W1259-N27). W.\,N.\ acknowledges support from an ESQ fellowship of the Austrian Academy of Sciences (\"OAW). 

\appendix

\section{Noise correlation functions}\label{app_noise}

The correlation function of the noise terms $\xi_{\mathrm{\alpha}}$ and $\xi_{\mathrm{p}}$ represented in Eq.~\eqref{Eq.CoupledDGL}, and their cross correlation are derived in detail in \cite{domokos2003mechanical} and read for the here chosen system as follows:
\begin{subequations}
\begin{align}
\left\langle \xi_{\mathrm{\alpha}}^{*}\xi_{\mathrm{\alpha}}\right\rangle &=\kappa+\Gamma_{0}\sum_{j}\cos^{2}\left(kx_{j}\right),\\
\left\langle \xi_{\mathrm{p}}\xi_{\mathrm{\alpha}}\right\rangle &=-i\hbar k\Gamma_{0}\alpha\sin\left(kx\right),\\
\left\langle \xi_{\mathrm{p}}^{*}\xi_{\mathrm{p}}\right\rangle &=2\hbar^{2}k^{2}\Gamma_{0}\left|\alpha\right|^{2}\left(\cos^{2}\left(kx\right)\bar{u}^{2}+\sin^{2}\left(kx\right)\right).
\end{align}
\end{subequations}
$\bar{u}$ is the averaged projection of the spontaneous emission direction onto the cavity axis. We chose $\bar{u}^{2}=2/5$.


\begin{thebibliography}{25}
\makeatletter
\providecommand \@ifxundefined [1]{
 \@ifx{#1\undefined}
}
\providecommand \@ifnum [1]{
 \ifnum #1\expandafter \@firstoftwo
 \else \expandafter \@secondoftwo
 \fi
}
\providecommand \@ifx [1]{
 \ifx #1\expandafter \@firstoftwo
 \else \expandafter \@secondoftwo
 \fi
}
\providecommand \natexlab [1]{#1}
\providecommand \enquote  [1]{#1}
\providecommand \bibnamefont  [1]{#1}
\providecommand \bibfnamefont [1]{#1}
\providecommand \citenamefont [1]{#1}
\providecommand \href@noop [0]{\@secondoftwo}
\providecommand \href [0]{\begingroup \@sanitize@url \@href}
\providecommand \@href[1]{\@@startlink{#1}\@@href}
\providecommand \@@href[1]{\endgroup#1\@@endlink}
\providecommand \@sanitize@url [0]{\catcode `\\12\catcode `\$12\catcode
  `\&12\catcode `\#12\catcode `\^12\catcode `\_12\catcode `\12\relax}
\providecommand \@@startlink[1]{}
\providecommand \@@endlink[0]{}
\providecommand \url  [0]{\begingroup\@sanitize@url \@url }
\providecommand \@url [1]{\endgroup\@href {#1}{\urlprefix }}
\providecommand \urlprefix  [0]{URL }
\providecommand \Eprint [0]{\href }
\providecommand \doibase [0]{http://dx.doi.org/}
\providecommand \selectlanguage [0]{\@gobble}
\providecommand \bibinfo  [0]{\@secondoftwo}
\providecommand \bibfield  [0]{\@secondoftwo}
\providecommand \translation [1]{[#1]}
\providecommand \BibitemOpen [0]{}
\providecommand \bibitemStop [0]{}
\providecommand \bibitemNoStop [0]{.\EOS\space}
\providecommand \EOS [0]{\spacefactor3000\relax}
\providecommand \BibitemShut  [1]{\csname bibitem#1\endcsname}
\let\auto@bib@innerbib\@empty

\bibitem [{\citenamefont {Horak}\ \emph {et~al.}(1997)\citenamefont {Horak},
  \citenamefont {Hechenblaikner}, \citenamefont {Gheri}, \citenamefont
  {Stecher},\ and\ \citenamefont {Ritsch}}]{Horak1997cavity}
  \BibitemOpen
  \bibfield  {author} {\bibinfo {author} {\bibfnamefont {P.}~\bibnamefont
  {Horak}}, \bibinfo {author} {\bibfnamefont {G.}~\bibnamefont
  {Hechenblaikner}}, \bibinfo {author} {\bibfnamefont {K.~M.}\ \bibnamefont
  {Gheri}}, \bibinfo {author} {\bibfnamefont {H.}~\bibnamefont {Stecher}}, \
  and\ \bibinfo {author} {\bibfnamefont {H.}~\bibnamefont {Ritsch}},\ }\enquote
  {\bibinfo {title} {Cavity-Induced Atom Cooling in the Strong Coupling
  Regime},}\ \href {\doibase 10.1103/PhysRevLett.79.4974} {\bibfield  {journal}
  {\bibinfo  {journal} {Phys. Rev. Lett.}\ }\textbf {\bibinfo {volume} {79}},\
  \bibinfo {pages} {4974} (\bibinfo {year} {1997})}\BibitemShut {NoStop}
\bibitem [{\citenamefont {Domokos}\ and\ \citenamefont
  {Ritsch}(2003)}]{domokos2003mechanical}
  \BibitemOpen
  \bibfield  {author} {\bibinfo {author} {\bibfnamefont {P.}~\bibnamefont
  {Domokos}}\ and\ \bibinfo {author} {\bibfnamefont {H.}~\bibnamefont
  {Ritsch}},\ }\enquote {\bibinfo {title} {Mechanical effects of light in
  optical resonators},}\ \href {\doibase 10.1364/JOSAB.20.001098} {\bibfield
  {journal} {\bibinfo  {journal} {J. Opt. Soc. Am. B}\ }\textbf {\bibinfo
  {volume} {20}},\ \bibinfo {pages} {1098} (\bibinfo {year}
  {2003})}\BibitemShut {NoStop}
\bibitem [{\citenamefont {J\"ager}\ \emph {et~al.}(2017)\citenamefont
  {J\"ager}, \citenamefont {Xu}, \citenamefont {Sch\"utz}, \citenamefont
  {Holland},\ and\ \citenamefont {Morigi}}]{jager2017semiclassical}
  \BibitemOpen
  \bibfield  {author} {\bibinfo {author} {\bibfnamefont {S.~B.}\ \bibnamefont
  {J\"ager}}, \bibinfo {author} {\bibfnamefont {M.}~\bibnamefont {Xu}},
  \bibinfo {author} {\bibfnamefont {S.}~\bibnamefont {Sch\"utz}}, \bibinfo
  {author} {\bibfnamefont {M.~J.}\ \bibnamefont {Holland}}, \ and\ \bibinfo
  {author} {\bibfnamefont {G.}~\bibnamefont {Morigi}},\ }\enquote {\bibinfo
  {title} {Semiclassical theory of synchronization-assisted cooling},}\ \href
  {\doibase 10.1103/PhysRevA.95.063852} {\bibfield  {journal} {\bibinfo
  {journal} {Phys. Rev. A}\ }\textbf {\bibinfo {volume} {95}},\ \bibinfo
  {pages} {063852} (\bibinfo {year} {2017})}\BibitemShut {NoStop}
\bibitem [{\citenamefont {Maunz}\ \emph {et~al.}(2004)\citenamefont {Maunz},
  \citenamefont {Puppe}, \citenamefont {Schuster}, \citenamefont {Syassen},
  \citenamefont {Pinkse},\ and\ \citenamefont {Rempe}}]{maunz2004cavity}
  \BibitemOpen
  \bibfield  {author} {\bibinfo {author} {\bibfnamefont {P.}~\bibnamefont
  {Maunz}}, \bibinfo {author} {\bibfnamefont {T.}~\bibnamefont {Puppe}},
  \bibinfo {author} {\bibfnamefont {I.}~\bibnamefont {Schuster}}, \bibinfo
  {author} {\bibfnamefont {N.}~\bibnamefont {Syassen}}, \bibinfo {author}
  {\bibfnamefont {P.~W.~H.}\ \bibnamefont {Pinkse}}, \ and\ \bibinfo {author}
  {\bibfnamefont {G.}~\bibnamefont {Rempe}},\ }\enquote {\bibinfo {title}
  {Cavity cooling of a single atom},}\ \href {\doibase 10.1038/nature02387}
  {\bibfield  {journal} {\bibinfo  {journal} {Nature}\ }\textbf {\bibinfo
  {volume} {428}},\ \bibinfo {pages} {50} (\bibinfo {year} {2004})}\BibitemShut
  {NoStop}
\bibitem [{\citenamefont {Murr}\ \emph {et~al.}(2006)\citenamefont {Murr},
  \citenamefont {Nu\ss{}mann}, \citenamefont {Puppe}, \citenamefont {Hijlkema},
  \citenamefont {Weber}, \citenamefont {Webster}, \citenamefont {Kuhn},\ and\
  \citenamefont {Rempe}}]{murr2006three}
  \BibitemOpen
  \bibfield  {author} {\bibinfo {author} {\bibfnamefont {K.}~\bibnamefont
  {Murr}}, \bibinfo {author} {\bibfnamefont {S.}~\bibnamefont {Nu\ss{}mann}},
  \bibinfo {author} {\bibfnamefont {T.}~\bibnamefont {Puppe}}, \bibinfo
  {author} {\bibfnamefont {M.}~\bibnamefont {Hijlkema}}, \bibinfo {author}
  {\bibfnamefont {B.}~\bibnamefont {Weber}}, \bibinfo {author} {\bibfnamefont
  {S.~C.}\ \bibnamefont {Webster}}, \bibinfo {author} {\bibfnamefont
  {A.}~\bibnamefont {Kuhn}}, \ and\ \bibinfo {author} {\bibfnamefont
  {G.}~\bibnamefont {Rempe}},\ }\enquote {\bibinfo {title} {Three-dimensional
  cavity cooling and trapping in an optical lattice},}\ \href {\doibase
  10.1103/PhysRevA.73.063415} {\bibfield  {journal} {\bibinfo  {journal} {Phys.
  Rev. A}\ }\textbf {\bibinfo {volume} {73}},\ \bibinfo {pages} {063415}
  (\bibinfo {year} {2006})}\BibitemShut {NoStop}
\bibitem [{\citenamefont {Schleier-Smith}\ \emph {et~al.}(2011)\citenamefont
  {Schleier-Smith}, \citenamefont {Leroux}, \citenamefont {Zhang},
  \citenamefont {Van~Camp},\ and\ \citenamefont {Vuleti\ifmmode~\acute{c}\else
  \'{c}\fi{}}}]{schleier2011optomechanical}
  \BibitemOpen
  \bibfield  {author} {\bibinfo {author} {\bibfnamefont {M.~H.}\ \bibnamefont
  {Schleier-Smith}}, \bibinfo {author} {\bibfnamefont {I.~D.}\ \bibnamefont
  {Leroux}}, \bibinfo {author} {\bibfnamefont {H.}~\bibnamefont {Zhang}},
  \bibinfo {author} {\bibfnamefont {M.~A.}\ \bibnamefont {Van~Camp}}, \ and\
  \bibinfo {author} {\bibfnamefont {V.}~\bibnamefont
  {Vuleti\ifmmode~\acute{c}\else \'{c}\fi{}}},\ }\enquote {\bibinfo {title}
  {Optomechanical Cavity Cooling of an Atomic Ensemble},}\ \href {\doibase
  10.1103/PhysRevLett.107.143005} {\bibfield  {journal} {\bibinfo  {journal}
  {Phys. Rev. Lett.}\ }\textbf {\bibinfo {volume} {107}},\ \bibinfo {pages}
  {143005} (\bibinfo {year} {2011})}\BibitemShut {NoStop}
\bibitem [{\citenamefont {Kiesel}\ \emph {et~al.}(2013)\citenamefont {Kiesel},
  \citenamefont {Blaser}, \citenamefont {Deli{\'c}}, \citenamefont {Grass},
  \citenamefont {Kaltenbaek},\ and\ \citenamefont
  {Aspelmeyer}}]{kiesel2013cavity}
  \BibitemOpen
  \bibfield  {author} {\bibinfo {author} {\bibfnamefont {N.}~\bibnamefont
  {Kiesel}}, \bibinfo {author} {\bibfnamefont {F.}~\bibnamefont {Blaser}},
  \bibinfo {author} {\bibfnamefont {U.}~\bibnamefont {Deli{\'c}}}, \bibinfo
  {author} {\bibfnamefont {D.}~\bibnamefont {Grass}}, \bibinfo {author}
  {\bibfnamefont {R.}~\bibnamefont {Kaltenbaek}}, \ and\ \bibinfo {author}
  {\bibfnamefont {M.}~\bibnamefont {Aspelmeyer}},\ }\enquote {\bibinfo {title}
  {Cavity cooling of an optically levitated submicron particle},}\ \href
  {\doibase 10.1073/pnas.1309167110} {\bibfield  {journal} {\bibinfo  {journal}
  {Proc. Natl. Acadl Sci. USA}\ }\textbf {\bibinfo {volume} {110}},\ \bibinfo
  {pages} {14180} (\bibinfo {year} {2013})}\BibitemShut {NoStop}
\bibitem [{\citenamefont {Millen}\ \emph {et~al.}(2015)\citenamefont {Millen},
  \citenamefont {Fonseca}, \citenamefont {Mavrogordatos}, \citenamefont
  {Monteiro},\ and\ \citenamefont {Barker}}]{millen2015cavity}
  \BibitemOpen
  \bibfield  {author} {\bibinfo {author} {\bibfnamefont {J.}~\bibnamefont
  {Millen}}, \bibinfo {author} {\bibfnamefont {P.~Z.~G.}\ \bibnamefont
  {Fonseca}}, \bibinfo {author} {\bibfnamefont {T.}~\bibnamefont
  {Mavrogordatos}}, \bibinfo {author} {\bibfnamefont {T.~S.}\ \bibnamefont
  {Monteiro}}, \ and\ \bibinfo {author} {\bibfnamefont {P.~F.}\ \bibnamefont
  {Barker}},\ }\enquote {\bibinfo {title} {Cavity Cooling a Single Charged
  Levitated Nanosphere},}\ \href {\doibase 10.1103/PhysRevLett.114.123602}
  {\bibfield  {journal} {\bibinfo  {journal} {Phys. Rev. Lett.}\ }\textbf
  {\bibinfo {volume} {114}},\ \bibinfo {pages} {123602} (\bibinfo {year}
  {2015})}\BibitemShut {NoStop}
\bibitem [{\citenamefont {Stickler}\ \emph {et~al.}(2016)\citenamefont
  {Stickler}, \citenamefont {Nimmrichter}, \citenamefont {Martinetz},
  \citenamefont {Kuhn}, \citenamefont {Arndt},\ and\ \citenamefont
  {Hornberger}}]{stickler2016rotranslational}
  \BibitemOpen
  \bibfield  {author} {\bibinfo {author} {\bibfnamefont {B.~A.}\ \bibnamefont
  {Stickler}}, \bibinfo {author} {\bibfnamefont {S.}~\bibnamefont
  {Nimmrichter}}, \bibinfo {author} {\bibfnamefont {L.}~\bibnamefont
  {Martinetz}}, \bibinfo {author} {\bibfnamefont {S.}~\bibnamefont {Kuhn}},
  \bibinfo {author} {\bibfnamefont {M.}~\bibnamefont {Arndt}}, \ and\ \bibinfo
  {author} {\bibfnamefont {K.}~\bibnamefont {Hornberger}},\ }\enquote {\bibinfo
  {title} {Rotranslational cavity cooling of dielectric rods and disks},}\
  \href {\doibase 10.1103/PhysRevA.94.033818} {\bibfield  {journal} {\bibinfo
  {journal} {Phys. Rev. A}\ }\textbf {\bibinfo {volume} {94}},\ \bibinfo
  {pages} {033818} (\bibinfo {year} {2016})}\BibitemShut {NoStop}
\bibitem [{\citenamefont {Fonseca}\ \emph {et~al.}(2016)\citenamefont
  {Fonseca}, \citenamefont {Aranas}, \citenamefont {Millen}, \citenamefont
  {Monteiro},\ and\ \citenamefont {Barker}}]{fonseca2016nonlinear}
  \BibitemOpen
  \bibfield  {author} {\bibinfo {author} {\bibfnamefont {P.~Z.~G.}\
  \bibnamefont {Fonseca}}, \bibinfo {author} {\bibfnamefont {E.~B.}\
  \bibnamefont {Aranas}}, \bibinfo {author} {\bibfnamefont {J.}~\bibnamefont
  {Millen}}, \bibinfo {author} {\bibfnamefont {T.~S.}\ \bibnamefont
  {Monteiro}}, \ and\ \bibinfo {author} {\bibfnamefont {P.~F.}\ \bibnamefont
  {Barker}},\ }\enquote {\bibinfo {title} {Nonlinear Dynamics and Strong Cavity
  Cooling of Levitated Nanoparticles},}\ \href {\doibase
  10.1103/PhysRevLett.117.173602} {\bibfield  {journal} {\bibinfo  {journal}
  {Phys. Rev. Lett.}\ }\textbf {\bibinfo {volume} {117}},\ \bibinfo {pages}
  {173602} (\bibinfo {year} {2016})}\BibitemShut {NoStop}
\bibitem [{\citenamefont {Windey}\ \emph {et~al.}(2018)\citenamefont {Windey},
  \citenamefont {Gonzalez-Ballestero}, \citenamefont {Maurer}, \citenamefont
  {Novotny}, \citenamefont {Romero-Isart},\ and\ \citenamefont
  {Reimann}}]{windey2018cavity}
  \BibitemOpen
  \bibfield  {author} {\bibinfo {author} {\bibfnamefont {D.}~\bibnamefont
  {Windey}}, \bibinfo {author} {\bibfnamefont {C.}~\bibnamefont
  {Gonzalez-Ballestero}}, \bibinfo {author} {\bibfnamefont {P.}~\bibnamefont
  {Maurer}}, \bibinfo {author} {\bibfnamefont {L.}~\bibnamefont {Novotny}},
  \bibinfo {author} {\bibfnamefont {O.}~\bibnamefont {Romero-Isart}}, \ and\
  \bibinfo {author} {\bibfnamefont {R.}~\bibnamefont {Reimann}},\ }\enquote
  {\bibinfo {title} {Cavity-Based 3D Cooling of a Levitated Nanoparticle via
  Coherent Scattering},}\ \href {https://arxiv.org/abs/1812.09176} {\bibfield
  {journal} {\bibinfo  {journal} {arXiv preprint arXiv:1812.09176}\ } (\bibinfo
  {year} {2018})}\BibitemShut {NoStop}
\bibitem [{\citenamefont {Magrini}\ \emph {et~al.}(2018)\citenamefont
  {Magrini}, \citenamefont {Norte}, \citenamefont {Riedinger}, \citenamefont
  {Marinkovi\'{c}}, \citenamefont {Grass}, \citenamefont {Deli\'{c}},
  \citenamefont {Gr\"{o}blacher}, \citenamefont {Hong},\ and\ \citenamefont
  {Aspelmeyer}}]{magrini2018near}
  \BibitemOpen
  \bibfield  {author} {\bibinfo {author} {\bibfnamefont {L.}~\bibnamefont
  {Magrini}}, \bibinfo {author} {\bibfnamefont {R.~A.}\ \bibnamefont {Norte}},
  \bibinfo {author} {\bibfnamefont {R.}~\bibnamefont {Riedinger}}, \bibinfo
  {author} {\bibfnamefont {I.}~\bibnamefont {Marinkovi\'{c}}}, \bibinfo
  {author} {\bibfnamefont {D.}~\bibnamefont {Grass}}, \bibinfo {author}
  {\bibfnamefont {U.}~\bibnamefont {Deli\'{c}}}, \bibinfo {author}
  {\bibfnamefont {S.}~\bibnamefont {Gr\"{o}blacher}}, \bibinfo {author}
  {\bibfnamefont {S.}~\bibnamefont {Hong}}, \ and\ \bibinfo {author}
  {\bibfnamefont {M.}~\bibnamefont {Aspelmeyer}},\ }\enquote {\bibinfo {title}
  {Near-field coupling of a levitated nanoparticle to a photonic crystal
  cavity},}\ \href {\doibase 10.1364/OPTICA.5.001597} {\bibfield  {journal}
  {\bibinfo  {journal} {Optica}\ }\textbf {\bibinfo {volume} {5}},\ \bibinfo
  {pages} {1597} (\bibinfo {year} {2018})}\BibitemShut {NoStop}
\bibitem [{\citenamefont {Wolke}\ \emph {et~al.}(2012)\citenamefont {Wolke},
  \citenamefont {Klinner}, \citenamefont {Ke{\ss}ler},\ and\ \citenamefont
  {Hemmerich}}]{wolke2012cavity}
  \BibitemOpen
  \bibfield  {author} {\bibinfo {author} {\bibfnamefont {M.}~\bibnamefont
  {Wolke}}, \bibinfo {author} {\bibfnamefont {J.}~\bibnamefont {Klinner}},
  \bibinfo {author} {\bibfnamefont {H.}~\bibnamefont {Ke{\ss}ler}}, \ and\
  \bibinfo {author} {\bibfnamefont {A.}~\bibnamefont {Hemmerich}},\ }\enquote
  {\bibinfo {title} {Cavity Cooling Below the Recoil Limit},}\ \href {\doibase
  10.1126/science.1219166} {\bibfield  {journal} {\bibinfo  {journal}
  {Science}\ }\textbf {\bibinfo {volume} {337}},\ \bibinfo {pages} {75}
  (\bibinfo {year} {2012})}\BibitemShut {NoStop}
\bibitem [{\citenamefont {Gangl}\ and\ \citenamefont
  {Ritsch}(2000)}]{gangl20003d}
  \BibitemOpen
  \bibfield  {author} {\bibinfo {author} {\bibfnamefont {M.}~\bibnamefont
  {Gangl}}\ and\ \bibinfo {author} {\bibfnamefont {H.}~\bibnamefont {Ritsch}},\
  }\enquote {\bibinfo {title} {3D dissipative motion of atoms in a strongly
  coupled driven cavity},}\ \href {\doibase 10.1007/s10053-000-9064-x}
  {\bibfield  {journal} {\bibinfo  {journal} {Eur. Phys. J. D}\ }\textbf
  {\bibinfo {volume} {8}},\ \bibinfo {pages} {29} (\bibinfo {year}
  {2000})}\BibitemShut {NoStop}
\bibitem [{\citenamefont {Neumeier}\ \emph {et~al.}(2018)\citenamefont
  {Neumeier}, \citenamefont {Northup},\ and\ \citenamefont
  {Chang}}]{neumeier2018reaching}
  \BibitemOpen
  \bibfield  {author} {\bibinfo {author} {\bibfnamefont {L.}~\bibnamefont
  {Neumeier}}, \bibinfo {author} {\bibfnamefont {T.~E.}\ \bibnamefont
  {Northup}}, \ and\ \bibinfo {author} {\bibfnamefont {D.~E.}\ \bibnamefont
  {Chang}},\ }\enquote {\bibinfo {title} {Reaching the optomechanical
  strong-coupling regime with a single atom in a cavity},}\ \href {\doibase
  10.1103/PhysRevA.97.063857} {\bibfield  {journal} {\bibinfo  {journal} {Phys.
  Rev. A}\ }\textbf {\bibinfo {volume} {97}},\ \bibinfo {pages} {063857}
  (\bibinfo {year} {2018})}\BibitemShut {NoStop}
\bibitem [{\citenamefont {Walls}\ and\ \citenamefont
  {Milburn}(1994)}]{wallsbook}
  \BibitemOpen
  \bibfield  {author} {\bibinfo {author} {\bibfnamefont {D.~F.}\ \bibnamefont
  {Walls}}\ and\ \bibinfo {author} {\bibfnamefont {G.~J.}\ \bibnamefont
  {Milburn}},\ }\href@noop {} {\emph {\bibinfo {title} {Quantum Optics}}},\
  \bibinfo {edition} {1st}\ ed.\ (\bibinfo  {publisher} {Springer-Verlag},\
  \bibinfo {address} {Berlin},\ \bibinfo {year} {1994})\BibitemShut {NoStop}
\bibitem [{\citenamefont {Ritsch}\ \emph {et~al.}(2013)\citenamefont {Ritsch},
  \citenamefont {Domokos}, \citenamefont {Brennecke},\ and\ \citenamefont
  {Esslinger}}]{ritsch2013cold}
  \BibitemOpen
  \bibfield  {author} {\bibinfo {author} {\bibfnamefont {H.}~\bibnamefont
  {Ritsch}}, \bibinfo {author} {\bibfnamefont {P.}~\bibnamefont {Domokos}},
  \bibinfo {author} {\bibfnamefont {F.}~\bibnamefont {Brennecke}}, \ and\
  \bibinfo {author} {\bibfnamefont {T.}~\bibnamefont {Esslinger}},\ }\enquote
  {\bibinfo {title} {Cold atoms in cavity-generated dynamical optical
  potentials},}\ \href {\doibase 10.1103/RevModPhys.85.553} {\bibfield
  {journal} {\bibinfo  {journal} {Rev. Mod. Phys.}\ }\textbf {\bibinfo {volume}
  {85}},\ \bibinfo {pages} {553} (\bibinfo {year} {2013})}\BibitemShut
  {NoStop}
\bibitem [{\citenamefont {Gardiner}\ and\ \citenamefont
  {Zoller}(2000)}]{gardinerbook}
  \BibitemOpen
  \bibfield  {author} {\bibinfo {author} {\bibfnamefont {C.~W.}\ \bibnamefont
  {Gardiner}}\ and\ \bibinfo {author} {\bibfnamefont {P.}~\bibnamefont
  {Zoller}},\ }\href@noop {} {\emph {\bibinfo {title} {Quantum Noise}}},\
  \bibinfo {edition} {2nd}\ ed.\ (\bibinfo  {publisher} {Springer-Verlag},\
  \bibinfo {address} {Berlin},\ \bibinfo {year} {2000})\BibitemShut {NoStop}
\bibitem [{\citenamefont {Maschler}\ and\ \citenamefont
  {Ritsch}(2005)}]{maschler2005cold}
  \BibitemOpen
  \bibfield  {author} {\bibinfo {author} {\bibfnamefont {C.}~\bibnamefont
  {Maschler}}\ and\ \bibinfo {author} {\bibfnamefont {H.}~\bibnamefont
  {Ritsch}},\ }\enquote {\bibinfo {title} {Cold Atom Dynamics in a Quantum
  Optical Lattice Potential},}\ \href {\doibase 10.1103/PhysRevLett.95.260401}
  {\bibfield  {journal} {\bibinfo  {journal} {Phys. Rev. Lett.}\ }\textbf
  {\bibinfo {volume} {95}},\ \bibinfo {pages} {260401} (\bibinfo {year}
  {2005})}\BibitemShut {NoStop}
\bibitem [{\citenamefont {Puppe}\ \emph {et~al.}(2007)\citenamefont {Puppe},
  \citenamefont {Schuster}, \citenamefont {Grothe}, \citenamefont {Kubanek},
  \citenamefont {Murr}, \citenamefont {Pinkse},\ and\ \citenamefont
  {Rempe}}]{puppe2007trapping}
  \BibitemOpen
  \bibfield  {author} {\bibinfo {author} {\bibfnamefont {T.}~\bibnamefont
  {Puppe}}, \bibinfo {author} {\bibfnamefont {I.}~\bibnamefont {Schuster}},
  \bibinfo {author} {\bibfnamefont {A.}~\bibnamefont {Grothe}}, \bibinfo
  {author} {\bibfnamefont {A.}~\bibnamefont {Kubanek}}, \bibinfo {author}
  {\bibfnamefont {K.}~\bibnamefont {Murr}}, \bibinfo {author} {\bibfnamefont
  {P.~W.~H.}\ \bibnamefont {Pinkse}}, \ and\ \bibinfo {author} {\bibfnamefont
  {G.}~\bibnamefont {Rempe}},\ }\enquote {\bibinfo {title} {Trapping and
  Observing Single Atoms in a Blue-Detuned Intracavity Dipole Trap},}\ \href
  {\doibase 10.1103/PhysRevLett.99.013002} {\bibfield  {journal} {\bibinfo
  {journal} {Phys. Rev. Lett.}\ }\textbf {\bibinfo {volume} {99}},\ \bibinfo
  {pages} {013002} (\bibinfo {year} {2007})}\BibitemShut {NoStop}
\bibitem [{\citenamefont {Hechenblaikner}\ \emph {et~al.}(1998)\citenamefont
  {Hechenblaikner}, \citenamefont {Gangl}, \citenamefont {Horak},\ and\
  \citenamefont {Ritsch}}]{hechenblaikner1998cooling}
  \BibitemOpen
  \bibfield  {author} {\bibinfo {author} {\bibfnamefont {G.}~\bibnamefont
  {Hechenblaikner}}, \bibinfo {author} {\bibfnamefont {M.}~\bibnamefont
  {Gangl}}, \bibinfo {author} {\bibfnamefont {P.}~\bibnamefont {Horak}}, \ and\
  \bibinfo {author} {\bibfnamefont {H.}~\bibnamefont {Ritsch}},\ }\enquote
  {\bibinfo {title} {Cooling an atom in a weakly driven high-$Q$ cavity},}\
  \href {\doibase 10.1103/PhysRevA.58.3030} {\bibfield  {journal} {\bibinfo
  {journal} {Phys. Rev. A}\ }\textbf {\bibinfo {volume} {58}},\ \bibinfo
  {pages} {3030} (\bibinfo {year} {1998})}\BibitemShut {NoStop}
\bibitem [{\citenamefont {Domokos}\ \emph {et~al.}(2002)\citenamefont
  {Domokos}, \citenamefont {Salzburger},\ and\ \citenamefont
  {Ritsch}}]{domokos2002dissipative}
  \BibitemOpen
  \bibfield  {author} {\bibinfo {author} {\bibfnamefont {P.}~\bibnamefont
  {Domokos}}, \bibinfo {author} {\bibfnamefont {T.}~\bibnamefont {Salzburger}},
  \ and\ \bibinfo {author} {\bibfnamefont {H.}~\bibnamefont {Ritsch}},\
  }\enquote {\bibinfo {title} {Dissipative motion of an atom with transverse
  coherent driving in a cavity with many degenerate modes},}\ \href {\doibase
  10.1103/PhysRevA.66.043406} {\bibfield  {journal} {\bibinfo  {journal} {Phys.
  Rev. A}\ }\textbf {\bibinfo {volume} {66}},\ \bibinfo {pages} {043406}
  (\bibinfo {year} {2002})}\BibitemShut {NoStop}
\bibitem [{\citenamefont {Niedenzu}\ \emph {et~al.}(2011)\citenamefont
  {Niedenzu}, \citenamefont {Grie{\ss}er},\ and\ \citenamefont
  {Ritsch}}]{niedenzu2011kinetic}
  \BibitemOpen
  \bibfield  {author} {\bibinfo {author} {\bibfnamefont {W.}~\bibnamefont
  {Niedenzu}}, \bibinfo {author} {\bibfnamefont {T.}~\bibnamefont
  {Grie{\ss}er}}, \ and\ \bibinfo {author} {\bibfnamefont {H.}~\bibnamefont
  {Ritsch}},\ }\enquote {\bibinfo {title} {Kinetic theory of cavity cooling and
  self-organisation of a cold gas},}\ \href {\doibase
  10.1209/0295-5075/96/43001} {\bibfield  {journal} {\bibinfo  {journal} {EPL}\
  }\textbf {\bibinfo {volume} {96}},\ \bibinfo {pages} {43001} (\bibinfo {year}
  {2011})}\BibitemShut {NoStop}
\bibitem [{\citenamefont {Asb\'oth}\ \emph {et~al.}(2005)\citenamefont
  {Asb\'oth}, \citenamefont {Domokos}, \citenamefont {Ritsch},\ and\
  \citenamefont {Vukics}}]{asboth2005self}
  \BibitemOpen
  \bibfield  {author} {\bibinfo {author} {\bibfnamefont {J.~K.}\ \bibnamefont
  {Asb\'oth}}, \bibinfo {author} {\bibfnamefont {P.}~\bibnamefont {Domokos}},
  \bibinfo {author} {\bibfnamefont {H.}~\bibnamefont {Ritsch}}, \ and\ \bibinfo
  {author} {\bibfnamefont {A.}~\bibnamefont {Vukics}},\ }\enquote {\bibinfo
  {title} {Self-organization of atoms in a cavity field: Threshold,
  bistability, and scaling laws},}\ \href {\doibase 10.1103/PhysRevA.72.053417}
  {\bibfield  {journal} {\bibinfo  {journal} {Phys. Rev. A}\ }\textbf {\bibinfo
  {volume} {72}},\ \bibinfo {pages} {053417} (\bibinfo {year}
  {2005})}\BibitemShut {NoStop}
\bibitem [{\citenamefont {Piazza}\ and\ \citenamefont
  {Ritsch}(2015)}]{piazza2015self}
  \BibitemOpen
  \bibfield  {author} {\bibinfo {author} {\bibfnamefont {F.}~\bibnamefont
  {Piazza}}\ and\ \bibinfo {author} {\bibfnamefont {H.}~\bibnamefont
  {Ritsch}},\ }\enquote {\bibinfo {title} {Self-Ordered Limit Cycles, Chaos,
  and Phase Slippage with a Superfluid inside an Optical Resonator},}\ \href
  {\doibase 10.1103/PhysRevLett.115.163601} {\bibfield  {journal} {\bibinfo
  {journal} {Phys. Rev. Lett.}\ }\textbf {\bibinfo {volume} {115}},\ \bibinfo
  {pages} {163601} (\bibinfo {year} {2015})}\BibitemShut {NoStop}
\end{thebibliography}
\end{document}